\documentclass[12pt]{article}
\usepackage{amssymb}
\usepackage{amsmath}
\usepackage[all]{xy}
\newcommand{\ft}[2]{{\textstyle\frac{#1}{#2}}}

\def\Tr{\mathop{\rm Tr}\nolimits}
\def\tr{\mathop{\rm tr}\nolimits}
\def\rme{{\mathrm e}}
\def\rmi{{\mathrm i}}
\newcommand{\ii}{\mathrm{i}}
\def\rmd{{\mathrm d}}
\newsavebox{\uuunit}
\sbox{\uuunit}
    {\setlength{\unitlength}{0.825em}
     \begin{picture}(0.6,0.7)
        \thinlines
        \put(0,0){\line(1,0){0.5}}
        \put(0.15,0){\line(0,1){0.7}}
        \put(0.35,0){\line(0,1){0.8}}
       \multiput(0.3,0.8)(-0.04,-0.02){12}{\rule{0.5pt}{0.5pt}}
     \end {picture}}
\newcommand {\unity}{\mathord{\!\usebox{\uuunit}}}

\csname @addtoreset\endcsname{equation}{section}
\newcommand{\SU}{\mathop{\rm SU}}
\newcommand{\SO}{\mathop{\rm SO}}
\newcommand{\U}{\mathop{\rm {}U}}
\newcommand{\USp}{\mathop{\rm {}USp}}

\newcommand{\Sp}{\mathop{\rm {}Sp}}

\newcommand{\fsu}{\mathfrak{su}}
\newcommand{\fsp}{\mathfrak{sp}}
\newcommand{\fusp}{\mathfrak{usp}}

\newcommand{\rSU}{\mathrm{SU}}
\newcommand{\rU}{\mathrm{U}}
\newcommand{\VV}{{\cal V}}

\def\cL{{\cal L}}
\def\cN{{\cal N}}
\def\cP{{\cal P}}

\newtheorem{note}{Remark}[section]
\newtheorem{lemma}{Lemma}[section]
\newtheorem{corollary}{Corollary}[section]
\newtheorem{proposition}{Proposition}[section]

\newcommand{\upomega}{\mbox{\usefont{U}{psy}{m}{n}w}}

\begin{document}
 %%%%%%%%%%%%%%%%%%%%%%%%%%%%%%%%%%%%%%%%%%%%%%%%%%%%%%%%%%%
\begin{titlepage}
\begin{flushright}
UB-ECM-PF 06/29\\
hep-th/0610300
\end{flushright}
\vspace{.5cm}
\begin{center}
\baselineskip=16pt
{\LARGE  Null Deformed Domain Wall  % \\ \vskip 0.2cm between lines
}\\
\vfill%\vskip 15mm%27.mm
{\large Alessio Celi
} \\
\vfill%\vskip 7mm%1cm
{\small Departament ECM, Facultat de F\'isica, Universitat de Barcelona, \\
Diagonal 647, E-08028 Barcelona, Spain.\\
 }
\end{center}
\vfill
\begin{center}
{\bf Abstract}
\end{center}
{\small We study null 1/4 BPS deformations of flat domain wall solutions (NDDW) in $\cN=2$, $d=5$ gauged
supergravity with hypermultiplets and vector multiplets  coupled. These are uncharged {\it time-dependent}
configurations and contain as special case, 1/2 supersymmetric flat domain walls (DW), as well as 1/2 BPS null
solutions of the ungauged supergravity. Combining our analysis with the classification method initiated by {\it
Gauntlett et al.}, we prove that {\it all the possible deformations of the DW have origin in the hypermultiplet
sector} or/and {\it are null}. Here, we classify all the null deformations: we show that they naturally organize
themselves into ``gauging" ($v$-deformation) and ``non gauging" ($u$-deformation). They have different
properties: only in presence of $v$-deformation is the solution supported by a {\it time-dependent} scalar
potential. Furthermore we show that the number of possible deformations {\it equals} the number of matter
multiplets coupled. We discuss the general procedure for constructing explicit solutions, stressing the crucial
role taken by the integrability conditions of the scalars as spacetime functions. Two analytical solutions are
presented. Finally, we comment on the holographic applications of the NDDW, in relation to the recently proposed
{\it time-dependent AdS/CFT}.}\vspace{2mm} \vfill \hrule width 3.cm

\vspace{3mm}
 {\footnotesize \noindent
e-mail: alessio@ecm.ub.es}
\end{titlepage}
\addtocounter{page}{1}
 \tableofcontents{}
\newpage
%%%%%%%%%%%%%%%%

\section{Introduction}

Studying time-dependent solutions in (Super)Gravity and String theory is an interesting and difficult task.
Indeed our capacity of producing  efficient cosmological models and generally describing our world relies on our
control over time evolution.  String theory, as a consistent theory of quantum gravity, should be able to
provide a satisfactory answer to this and other outstanding related problems, such as the resolution of
spacetime singularities. Unfortunately, it is very hard to keep the stability of such solutions under control,
especially against quantum corrections. One of the crucial points is that a generic time-dependent solution is
{\it not supersymmetric}, thus does not enjoy non renormalization properties associated to BPS configurations.
Up to now the use of this property is the main way we have to study non perturbative phenomena.

In this work we take the modest approach of considering an interesting class of time-dependent BPS
configurations in $\cN=2$ $d=5$ gauged supergravity with matter couplings. In doing so, we are in part inspired
by \cite{Chu:2006pa,Das:2006dz,Lin:2006ie},\footnote{In \cite{Lin:2006ie} null deformations of the near horizon
limit of the general D$p$-branes are considered and their holographic properties are studied. For $p\neq3$, they
are the null generalization of the $(p+2)$-dimensional domain walls of \cite{Boonstra:1998mp}.} where null
deformations of $AdS_5\times S^5$ are considered. These authors propose an extension of the AdS/CFT
correspondence to such background, that is the near horizon limit of a null deformed stack of D3-branes (the
null deformation of intersecting brane configurations has been recently considered in \cite{Ohta:2006sw}). Such
an extension is appealing because may allow one to inspect toy spacetime cosmological singularities via
holography. In \cite{Chu:2006pa,Das:2006dz,Lin:2006ie} it is argued that the dual theory corresponds to $\cN=4$
super Yang--Mills theory (SYM)
 with time-dependent sources turned on. This picture has been supported and further
investigated in \cite{Das:2006pw}. An interesting property of the background analyzed in
\cite{Chu:2006pa,Das:2006dz,Lin:2006ie} is that the dilaton and, consequently, the gauge coupling of the dual
theory are time-dependent (through a lightcone coordinate). Furthermore, as only the $AdS_5$ part is affected by
the deformation, such solutions can be studied in full generality in the effective 5d (gauged) supergravity.

In our paper we investigate configurations of the form
\begin{equation*}
\rmd s^2 = \beta^2(x^+, r)\left(-2k^2(x^+) \rmd x^+\rmd x^- + H(x^+,x^-,x^i,r) (\rmd x^+)^2 + (\rmd x^i)^2  +
\rmd r^2\right).
\end{equation*}
We show that such configurations preserve 1/4-supersymmetry and include the null deformed $AdS_5$ space of
\cite{Chu:2006pa,Das:2006dz,Lin:2006ie} as special 1/2 BPS subcases. However, the above metric describes also
another interesting 1/2-supersymmetric subclass - it contains flat domain wall solutions. This class of
solutions has received a lot of attention mainly due to the role in the AdS/CFT correspondence,
\cite{Boonstra:1998mp}. As solutions of gauged supergravity these are conjectured to be dual to the
Renormalization Group (RG) flows of field theory couplings
\cite{Girardello:1999bd,Girardello:1998pd,Girardello:1999hj,Gubser:1999eu,DeWolfe:1999cp,Skenderis:1999mm,Klebanov:2000nc}.
Domain walls are also a key ingredient of Brane world constructions
\cite{Horava:1995qa,Horava:1996ma,Lukas:1998yy,Randall:1999vf,Lykken:1999nb,Randall:1999ee}.
 More recently, in four dimensions, these solitons have been used as a laboratory for understanding mirror
symmetry in flux/generalized geometry compactifications \cite{Mayer:2004sd,House:2004hv,Louis:2006wq} and to
explore transitions between the different cosmological vacua of the Landscape \cite{Ceresole:2006iq}.

It is desirable to ``combine" the two deformations of AdS we consider  and verify whether and/or how the
gauge/gravity correspondence applies to the resulting background. For these reasons, the study of ``generalized"
domain wall solutions remains an interesting area of study. Very recently non-supersymmetric charged domain
walls have been investigated in \cite{Gutowski:2006hk} while BPS {\it gyratons} have been discussed in
\cite{Caldarelli:2006ww}. In both cases such configurations have been studied in the presence of vector
multiplets coupling only.

In this work we shall consider all the matter couplings that are relevant for constructing domain walls. As
shown in \cite{Ceresole:2001wi} the inclusion of hypermultiplets is crucial to have BPS domain walls
interpolating between two AdS vacua and consequently to embed the domain wall solution of \cite{Freedman:1999gp}
(FGPW) in the $\cN=2$ gauged supergravity, as holographic dual to an RG flow from an $\cN = 4$ to an $\cN = 1$
SYM. In \cite{Celi:2004st} it has been shown that curved domain walls can be obtained only with hypermultiplets
coupled. Currently there is a renewed interest in having a more systematical understanding of BPS solutions with
hypermultiplets. The full classification in $\cN=2$ ungauged supergravity has been achieved in four and five
dimensions in \cite{Hubscher:2006mr,Bellorin:2006yr}. Some steps towards this goal in the more complicated
gauged case had been previously performed in five dimensions
\cite{Cacciatori:2002qx,Celi:2003qk,Cacciatori:2004qm}.

The configurations we present here are the first example of BPS time-dependent solutions in gauged supergravity
with hypermultiplet coupled.

As an additional motivation, we would like to mention that the configurations we consider may be seen as the
closest supersymmetry-preserving analogue of time-dependent solutions of \cite{Gibbons:2005rt,Chen:2005jp}
describing Brane collision.

The organization of the paper is as follows. In order to fix the notation and be self-contained we present in
section 2 the basic ingredients of the supergravity theory we are dealing with and we describe the main feature
of (flat) domain wall solutions in $\cN=2$ $d=5$ gauged supergravity.

Section \ref{null} constitutes the main part of this paper and is devoted to the derivation  and discussion of
the BPS equations related to the metric above. Such an analysis is made in comparison with the original domain
wall case which, using a non orthodox English terminology, we will refer to in the text as the ``undeformed"
configuration. We shall illustrate how the class of solitons under consideration admits a dual interpretation as
null deformation of domain walls or deformation of a plane wave due to the ``gauging". Taking the first point of
view, we show that the null deformation naturally organizes itself into the contribution coming from the
gauging, and another associated to the null solutions in the ungauged supergravity.

In section \ref{sol} the analysis of section \ref{null} is given concrete applications and two explicit examples
are constructed.

We finally collect our conclusions and propose possible developments in section \ref{conclu}.

All details of calculation that have not been given in the main text are presented in the Appendix \ref{MIC} and
\ref{eom}. In appendix \ref{app:paramCosets} we describe the parametrization of the coset space that appears in
section \ref{solh}. In appendix \ref{adapt} we argue how ``adapted coordinates" can be used to derive some
insights into the possible solutions.

\section{Domain wall in $\cN=2$ $d=5$ gauged supergravity}

This section is devoted mainly to review known facts on domain wall solution. Furthermore we remind here the
basic ingredient of the supergravity theory we use, giving the formulae we use in our calculation.

\subsection{Five-dimensional, $\mathcal{N}=2$ gauged supergravity}
\label{ss:5dN2SG}

We start by  recalling some of the  most important   features of five-dimensional, $\mathcal{N}=2$ gauged
supergravity theories. Further technical details can be found in the original references
\cite{Gunaydin:1984bi,Gunaydin:1985ak,Gunaydin:1999zx,Ceresole:2000jd,Bergshoeff:2004kh}.

The matter multiplets that can be coupled to $5D$, $\mathcal{N}=2$ supergravity are vector, tensor and
hypermultiplets: the scalar $\varphi$ of theory could a priori sit in any of these (or even be a combination of
different types of scalars).

The $(n_V+n_T)$ scalar fields of $n_{V}$ vector and  $n_T$ tensor multiplets parameterize a ``very special''
real  manifold $\mathcal{M}_{\rm VS}$, i.e., an $\left(n_{V}+n_{T}\right)$--dimensional hypersurface of an
auxiliary $(n_{V}+n_{T}+1)$-dimensional space spanned by coordinates $h^{\tilde{I}}$  $(\tilde{I}=0,1,\ldots,
n_{V}+n_{T}+1)$ :
\begin{equation}
\mathcal{M}_{\rm VS}=\{ h^{\tilde{I}}\in \mathbb{R}^{(n_V+n_T+1)}:
C_{\tilde{I}\tilde{J}\tilde{K}}h^{\tilde{I}}h^{\tilde{J}}h^{\tilde{K}}=1\}, \label{defVS}
\end{equation}
where the constants $C_{\tilde{I}\tilde{J}\tilde{K}}$ appear in a Chern-Simons-type coupling
 of the Lagrangian.
The embedding coordinates $h^{\tilde{I}}$
 have a natural splitting,
\begin{equation}
h^{\tilde{I}}=(h^I,h^M), \qquad (I=0,1,\ldots,n_{V}), \qquad (M=1,\ldots, n_{T}),
\end{equation}
where the $h^{I}$ are related to the sub-geometry of the $n_{V}$ vector multiplets, and the $h^{M}$ refer to the
$n_{T}$ tensor multiplets. On $\mathcal{M}_{\rm VS}$, the $h^{\tilde{I}}$ become functions of the physical
scalar fields, $\phi^{x}$ $(x=1,\ldots,n_{V}+n_{T})$. The metric on the very special manifold is determined via
the equations
\begin{eqnarray}
  &&g_{xy}=h_x^{\tilde{I}}\,h_{y\tilde{I}},\qquad h_x^{\tilde{I}}\equiv
  -\sqrt{\ft32}\,\partial _xh^{\tilde{I}}, \qquad
  h_{\tilde{I}}\equiv C_{\tilde{I}\tilde{J}\tilde{K}}h^{\tilde{J}}h^{\tilde{K}},
\qquad h_{\tilde{I}x}\equiv \sqrt{\ft32}\,\partial _xh_{\tilde{I}},\nonumber\\
  &&h^{\tilde{I}}h_{\tilde{J}}+h_x^{\tilde{I}}\,g^{xy}\,h_{y\tilde{J}}=
  \delta^{\tilde{I}}_{\tilde{J}}, \qquad h^{\tilde{I}}h_{\tilde{I}}=1,\qquad h^{\tilde{I}}h_{\tilde{I}x}=0.
\label{identVS}
\end{eqnarray}

The scalars  $q^X$ $(X=1,\ldots 4n_{H})$ of  $n_{H}$ hypermultiplets, on the other hand, take their values in a
quaternionic-K{\"a}hler manifold $\mathcal{M}_{\rm Q}$ \cite{Bagger:1983tt}, i.e., a manifold of real dimension
$4n_H$ with holonomy group contained in $SU(2)\times USp(2n_H)$. We denote the vielbein on this manifold by
$f_X^{iA}$, where $i=1,2$ and $A=1,\ldots,2n_{H}$ refer to an adapted $SU(2)\times USp(2n_{H})$ decomposition of
the tangent space. The hypercomplex structure is $(-2)$ times the curvature of the $SU(2)$ part of the holonomy
group\footnote{In fact, the proportionality factor includes the Planck mass and the metric, which are implicit
here.}, denoted as $\mathcal{R}^{rZX}$ $(r=1,2,3)$, so that the quaternionic identity reads
\begin{equation}
  {\cal R}^r_{XY}{\cal R}^{sYZ}=-\ft14\,\delta ^{rs}\,\delta _X{}^Z
   -\ft12\,\varepsilon ^{rst}\,{\cal R}^t_X{}^Z.
 \label{quaterid}
\end{equation}

Besides these scalar fields, the bosonic sector of the matter multiplets also contains $n_{T}$ tensor fields
$B^M_{\mu\nu}$ $(M=1,\ldots,n_{T})$ from the $n_{T}$ tensor multiplets and
 $n_{V}$  vector fields from the $n_{V}$ vector multiplets.
Including the graviphoton, we thus have a total of $(n_{V}+1)$ vector fields, $A_{\mu}^{I}$
$(I=0,1,\ldots,n_{V})$, which can be used to gauge up to $(n_{V}+1)$ isometries of the quaternionic-K{\"a}hler
manifold
$\mathcal{M}_{\rm Q}$ %and the very special manifold $\mathcal{M}_{VS}$
(provided such  isometries exist). These symmetries act on the vector-tensor multiplets by a representation
$t_{I\tilde{J}}{}^{\tilde K}$, where in the pure vector multiplet sector $t_{IJ}{}^K=f_{IJ}{}^K$ are the
structure constants, and the other components also satisfy  some restrictions
\cite{Gunaydin:1999zx,Ellis:2001xd,Bergshoeff:2004kh}. The transformations should leave the defining condition
in (\ref{defVS}) invariant, hence
\begin{equation}
  t_{I(\tilde{J}}{}^{\tilde M}C_{\tilde K\tilde L)\tilde M}=0.
 \label{restrgauging}
\end{equation}
The very special K{\"a}hler target space then has Killing vectors
\begin{equation}
  K^{x}_I(\phi)= -\sqrt{\ft32}t_{I\tilde{J}}{}^{\tilde K}h_{\tilde{K}}^{ x}h^{\tilde J}.
 \label{KillingVSM}
\end{equation}
There may be more Killing vectors, but these are the ones that are gauged using the gauge vectors in the vector
multiplets.

The quaternionic Killing vectors $K_{I}^{X}(q)$ that generate the isometries on $\mathcal{M}_{\rm Q}$ can be
expressed in terms of the derivatives of $SU(2)$ triplets of Killing prepotentials $P_{I}^{r}(q)$ $(r=1,2,3)$
via
\begin{equation}
D_{X}P^{r}_{I}= \mathcal{R}_{XY}^{r}K^{Y}_{I}, \qquad \Leftrightarrow \qquad \left\{ \begin{array}{c}
  K_{I}^Y=-\frac{4}{3}\mathcal{R}^{rYX}D_{X}P^{r}_{I} \\ [2mm]
  D_XP_I^r=-\varepsilon ^{rst}{\cal R}^s_{XY}D^YP_I^t,
\end{array}\right.
\label{KillingP}
\end{equation}
where $D_{X}$ denotes the $SU(2)$ covariant derivative, which contains an $SU(2)$ connection $\omega_{X}^{r}$
with curvature $ \mathcal{R}^r_{XY}$:
\begin{equation}
  D_X P^r= \partial _XP^r+2\,\varepsilon ^{rst} \omega _X^sP^t,\qquad
\mathcal{R}^r_{XY}=2\,\partial _{[X}\omega _{Y]}^r+2\,\varepsilon ^{rst}\omega _X^s\omega _Y^t.
 \label{defSU2cR}
\end{equation}
The prepotentials satisfy the constraint
\begin{equation}\label{constraint}
\frac{1}{2}\mathcal{R}_{XY}^{r}K_{I}^{X}K_{J}^{Y}- \varepsilon^{rst}P_{I}^{s}P_{J}^{t}
+\frac{1}{2}f_{IJ}{}^{K}P_{K}^{r}=0,
\end{equation}
where $f_{IJ}{}^{K}$ are the structure constants of the gauge group.

In the following, we will frequently switch between the above vector notation for $SU(2)$-valued quantities such
as $P_{I}^{r}$, and  the usual
   $(2\times 2)$   matrix notation,
\begin{equation}
%\mathbf{P}_I= \left(   P_{Ii}{}^j\right) ,\qquad
%  \equiv\rmi \vec \sigma_i{}^j\cdot\vec P_{I}
P_{Ii}{}^j\equiv \rmi\,\sigma_{ri}{}^jP_{I}^{r}  .
 \label{matrixVector}
\end{equation}

An important difference in geometrical significance between the very special Killing vectors $K_{I}^{x}(\phi)$
in  (\ref{KillingVSM})  and the quaternionic ones $K_{I}^{X}(q)$  in (\ref{KillingP}), is that the former do not
arise as derivatives of Killing prepotentials, because there is no natural symplectic structure on the real
manifold $\mathcal{M}_{\rm VS}$ that could define a moment map.\footnote{The moment maps are related to the fact
that the isometries should preserve complex structures. Therefore, they are absent in the real manifold. In 4
dimensions, the scalar manifold of the vector multiplets does have a complex structure. Hence, in that case this
sector would also have a moment map structure~\cite{D'Auria:1991fj}. This suggests that in four dimensions the
same comparison may go along different lines.}

Turning on only the metric and the scalars, the general Lagrangian of such a gauged supergravity  theory is

\begin{equation}
e^{-1}\mathcal{L}= \frac{1}{2}R-\frac{1}{2}g_{xy}\partial_{\mu}\phi^{x}
\partial^{\mu}\phi^{y}-\frac{1}{2}g_{XY}\partial_{\mu}q^{X}
\partial^{\mu}q^{Y}-g^2\mathcal{V}(\phi,q), \label{lagra}
\end{equation}
whereas the supersymmetry transformation laws of the fermions are given by
\begin{eqnarray}
\delta \psi_{\mu i}&=& {\nabla}_\mu\epsilon_i-  \omega_{\mu i}{}^{j} \epsilon_{j}- \frac{\rmi}{\sqrt{6}}\,
g\,\gamma_{\mu}P_{i}^{\,\, j}\epsilon_{j},\label{gravitino}\\
\delta \lambda_{i}^{x}&=& -\frac{\rmi}{2}\gamma^{\mu}(\partial_{\mu}\phi^{x})\epsilon_{i}
-g\,P_{i}{}^{jx}\epsilon_{j}+g\,\mathcal{T}^x\epsilon_{i},\label{gaugino}\\
\delta \zeta^{A}&=&\frac{\rmi}{2}f_{X}^{iA}\gamma^{\mu}(\partial_{\mu}q^{X}) \epsilon_{i} - g\,
\mathcal{N}^{iA}\epsilon_{i}.\label{hyperino}
\end{eqnarray}
Here, $\psi_{\mu}^{i}$, $\lambda_{i}^{x}$, $\zeta^{A}$ are the gravitini, gaugini (tensorini) and hyperini,
respectively,
 $g$ denotes the gauge coupling, the $SU(2)$ connection $\upomega_{\mu}$ is
defined as $\omega_{\mu i}{}^{j}= (\partial_{\mu}q^{X}) \omega_{X i}{}^{j} $, and
\begin{eqnarray}
P^{r}&=&h^{I}(\phi)P_{I}^{r}(q), \label{Pdefinition}  \\
P^{r}_{x}&=&-\sqrt{\frac{3}{2}}\partial_{x}P^r=h_{x}^{I}P_{I}^{r},\qquad P^{rx}=g^{xy}P^r_y,\\
\mathcal{N}^{iA}&=&\frac{\sqrt{6}}{4}f_X^{iA}(q)h^{I}(\phi)K_{I}^{X}(q),\label{hshift}\\
\mathcal{T}^{x}&=&\frac{\sqrt{6}}{4}h^{I}(\phi)K_{I}^{x}(\phi)
%=-\frac34t_{I\tilde J}{}^{\tilde K}h^Ih^{\tilde J}h^x_{\tilde K}
\label{tshift}.
\end{eqnarray}
As a general fact in supergravity, the potential is given by the sum of ``squares of the fermionic shifts'' (the
scalar expressions in the above transformations of the fermions):
\begin{equation}
\mathcal{V}=-4P^{r}P^{r} +2P_{x}^{r}P_{y}^{r}g^{xy}+2\mathcal{N}^{iA} \mathcal{N}^{jB}\varepsilon
_{ij}C_{AB}+2\mathcal{T}^x \mathcal{T}^yg_{xy}, \label{scalarpotential}
\end{equation}
where $C_{AB}$ is the (antisymmetric) symplectic metric of $USp(2n_{H})$.

Using the explicit form of the Killing vector, (\ref{KillingVSM}), in (\ref{tshift}), one finds that this
expression vanishes if the transformation matrix $t$ involves only vector multiplets. This is clear because then
$t_{IJ}{}^K=f_{IJ}{}^K$, hence antisymmetric. Therefore, the shift $\mathcal{T}^{x}$ in the above expressions is
non-vanishing only if there are charged tensor multiplets in the theory \footnote{In five dimensions, tensor
multiplets that are not charged under some gauge group are equivalent to vector multiplets. We always assume
that all uncharged tensor multiplets are converted to vector multiplets.}. Since $\mathcal{T}^{x}$ appears in
(\ref{gaugino}) with the unit matrix in $su(2)$ space, it must vanish on a BPS-domain wall solution for
compatibility with the spinor projector (see \cite[footnote 8]{Ceresole:2001wi} and \cite{Ceresole:2001zf}).
Furthermore, unlike the shifts $P_{x}^{r}$ and $\mathcal{N}^{iA}$, $\mathcal{T}^{x}$ is a purely ``D-type''
term, in the sense that it is completely unrelated to derivatives of the moment map $P^r$.
%Therefore, it can never fit the pattern (\ref{Freedman000}) of the fake supergravity transformations.
Thus, for BPS-domain walls in $5D$, $\mathcal{N}=2$ supergravity (and in fake supergravity as well
\cite{Celi:2004st}), non-trivial tensor multiplets can not play an important r\^{o}le, and we can limit our
remaining discussion to the case $n_T=0$, i.e., to supergravity coupled to vector and/or hypermultiplets only.
This also means that the index $\tilde{I}$ simply becomes the index $I$ in all previous equations, and the index
$M$ disappears.

Before reviewing the BPS  domain wall solutions, let us present the integrability conditions of the Gravitini
variation (\ref{gravitino}). Following \cite{Celi:2003qk}, all the information contained in (\ref{gravitino})
for uncharged BPS configurations in presence of matter, can be cast in the compact form:
\begin{eqnarray}
&& \left(\frac 14 {\Omega_{cd}}^{ab}\gamma_{ab} {\delta_i}^j - i R^r_{cd} {(\sigma_r)_i}^j - \frac{2g}{\sqrt 6 }
\gamma_{[c}D_{d]} P^r {(\sigma_r)_i}^j + \frac{g^2}{2} W^2 \gamma_{cd} \right) \epsilon_j = 0,\label{GIC}
\end{eqnarray}
where $D_\mu P^r \equiv  \partial_\mu \varphi^\Lambda D_\Lambda P^r$ and $R_{\mu\nu}^r \equiv \partial_\mu q^X
\partial_\nu q^Y R^r_{XY}$ are  the pull-back of the $\SU(2)$-covariant derivative of the moment map and of the
$\SU(2)$-curvature, respectively. $W$ is the {\it superpotential}, $P^rP^r \equiv \frac 32 W^2$ (the
normalization is chosen for convenience). Imposing (\ref{GIC}) together with the BPS conditions of the matter
field, is sufficient to ensure the Einstein equation for the metric, for time--like BPS configurations
\cite{Celi:2003qk} (i.e. when the vector bilinear constructed by the covariantly constant spinor $V^\mu \equiv
1/2\bar \epsilon^i\gamma^\mu\epsilon_i$ is time--like), or more precisely, when there is no light-like projector
on the covariantly constant spinor ($\gamma_\mp \epsilon = 0$). We will see in section \ref{null}, how the
equations of motion impose extra-condition over the metric in the light--like case ($V^\mu V_\mu =
0$).\footnote{Actually, a subtlety that has never been put in evidence is that a domain wall solution is always
light-like, although in a trivial way. This point will be clarified in section \ref{class}.}

\subsection{BPS-domain walls in supergravity}\label{DW}

Now we will remind to the reader of some known facts  about domain wall configurations, pointing out some novel
features along the way. This subject has been extensively studied in the literature, mainly as an
application/extention of the AdS/CFT correspondence and as phenomenological model with large extra-dimensions
(Brane world). The relevance of such configuration justified the derivation of an ``effective" supergravity
approach \cite{Skenderis:1999mm} known as  {\it Fake supergravity} \cite{Freedman:2003ax}, valid for any
space-time dimensions. The explicit relation of this powerful tool for constructing  domain wall solutions, with
the full-fledged $\cN=2$ $D=5$ gauged supergravity was first uncovered in \cite{Celi:2004st}, and further
explored in \cite{Skenderis:2006jq}.\footnote{The relation between Fake supergravity and $\cN=4$, $d=5$ gauged
supergravity has been studied in \cite{Zagermann:2004ac}.} Remarkably, the same first order formalism (extended
to include $dS$-brane in \cite{Afonso:2006gi,Bazeia:2006tj}) applies also to Friedmann-Robertson-Walker
cosmology \cite{Bazeia:2005tj}, motivating the derivation of the domain wall/Cosmology correspondence
\cite{Skenderis:2006jq,Skenderis:2006fb}.

 We will review the subject from a different prospective to usual (cfr. \cite{Celi:2004st}). The normal
 procedure is to start with a domain wall ansatz for the metric,
\begin{equation}
\rmd s^2=\rme^{2U(r)}g_{\bar \mu \bar \nu}(x)\,\rmd x^{\bar \mu} \,\rmd x^{\bar\nu} + \rmd r^2,
\label{curvedmetric}
\end{equation}
and assume that the scalar fields depend only on the fifth dimension $r$ (we indicate with a bar the indices
running over the remaining four dimensions). By definition of a domain wall, the four dimensional metric
$g_{\bar\mu\bar\nu}$ of the wall has constant curvature that BPS equations fixed to be non positive. When this
is negative ($AdS_4$) the domain wall is said to be {\it curve} or AdS-sliced, while is called {\it flat} or
Minkowski-sliced in case of zero curvature.

We shall instead begin by requiring that the scalar fields depending only on one spacetime  {\it spatial}
coordinate, that for convenience we take as fifth coordinate. This is equivalent to assume that the metric is a
{\it warped product} of a radial coordinate times a generic four dimensional metric. So any a priori assumption
is made about the form of $g_{\bar\mu\bar\nu}$ in (\ref{curvedmetric}),\footnote{Our study at this stage can not
exclude the existence of different supersymmetric solutions than the domain walls, where the scalars depend only
on one coordinate that does {\it not factorize } in the metric. Such possibility could be interesting in the
contest of holography.} a part the fact that it does not dependent on $r$. We will show of this weaker
requirement is sufficient to identify a domain wall solution. Following the analysis of \cite{Celi:2003qk},
further extended in \cite{hypereq}, we decompose the derivative of the quaternionic scalars as:
\begin{equation}
\partial_5q^X= M K^X + 2 v_{r} D^XP^r. \label{Dq}
\end{equation}
As a consequence, the hyperini equation (\ref{hyperino}) reduces to
\begin{eqnarray}
& &\left[\sqrt {\frac 32} i g {\delta_i}^j +\gamma_5 M {\delta_i}^j - i v^r \gamma_5
{(\sigma_r)_i}^j\right]\epsilon_j =0.\label{prov}
\end{eqnarray}
Now, the other crucial physical requirement of the solution enters the game, i.e. it must be {\it uncharged}.
Under this condition, the equations of motion for the gauge field reduces to
\begin{equation}
K^X\partial_a q^Y g_{XY} = 0,
\end{equation} which immediately gives $M=0$. Thus (\ref{prov}) becomes
\begin{equation}
\gamma_5 \epsilon_i = \alpha^r{(\sigma_r)_i}^j \epsilon_j,\label{PDW}
\end{equation}
where the phase $\alpha^r$ ($\alpha^r\alpha_r =1$) is given by $\alpha^r \equiv \sqrt{\frac 23} \frac 1g v^r$.

The analysis of the gaugini equation (\ref{gaugino}) yields to an analogous result. By imposing $\partial_a
\phi^x = \delta_a^5
\partial_5\phi^x$, one gets
\begin{equation}
\left(\partial_5 \phi^x\gamma_5{\delta_i}^j + 2g P^{xr} {(\sigma_r)_i}^j\right) \epsilon_j = 0.\label{prow}
\end{equation}
The above equation is easily seen to be equivalent to (\ref{PDW}) plus
\begin{equation}
\partial^x P^r = \frac {\alpha^r \partial_5 \phi^x}{\sqrt 6 g}.\label{dxP}
\end{equation}
Hence the first order equations for the scalars of hypermultiplets and vector multiplets can be written in a
unified framework as:
\begin{equation}
\partial_5 \varphi^\Lambda = 2 \sqrt{\frac 32} g \alpha^r D^\Lambda P^r, \ \ \ \ \  \Lambda= 1,\dots, n_V+4n_H,
\label{dotvarphi}
\end{equation}
where
$$\varphi^\Lambda \equiv  \begin{cases} q^X, & \Lambda = 1,\dots,4n_H = X\\
                                        \phi^x, &\Lambda = 4n_H+1,\dots,4n_H+n_V = x+ 4n_H
                          \end{cases},$$
$$D^\Lambda P^r           \begin{cases} D^X P^r, & \Lambda = 1,\dots,4n_H = X\\
                                        \partial^x P^r, &\Lambda = 4n_H+1,\dots,4n_H+n_V = x+ 4n_H
                          \end{cases},$$
However, let us emphasize that the vector multiplet scalar sector is constrained by a stronger condition, due to
(\ref{dxP}), i.e. $\partial_x P^r // \alpha^r$.\footnote{This property forces the domain wall supported by
vector multiplets to be flat, as first observed in \cite{Celi:2004st}.}

 We remember that, up to now, we did not assume any guess for the metric $g$ of the four dimensional slice orthogonal to
 $r$. Its form will be determined by the integrability conditions of the gravitini. Taking in account that, from
 (\ref{dotvarphi}) we find,
 \begin{eqnarray}
 D_a P^r &=& \partial_a \varphi^\Lambda D_\Lambda P^r = \cr
&=& 3 \sqrt{\frac{3}{2}} g \left(\partial^X W \partial_X W + \frac 1{\gamma^2} \partial^x W \partial_x W
\right)\delta_a^5 \alpha^r, \ \ \ \ \ \ \gamma \equiv - \alpha^s Q^s,
\end{eqnarray}
equation (\ref{GIC}) becomes
\begin{equation}
\left\{1/2 {\Omega_{cd}}^{ab}\gamma_{ab} - g^2 \left[3 \left(\partial^X W \partial_X W + \frac 1{\gamma^2}
\partial^x W \partial_x W \right) \left(\delta_c^5 + \delta_d^5\right)  - W^2 \right] \gamma_{cd} \right\}
\epsilon_i = 0,\label{intgradw}
\end{equation}
where (\ref{PDW})  is crucial to reduce the above expression to a combination of gamma matrices. Now,
differently from the case we will discuss in the next section, no other projection condition can be enforced
because we are looking for $1/2$ BPS solution. Hence, (\ref{intgradw}) must be trivial and the curvature
``diagonal", i.e.
\begin{equation}
{\Omega_{cd}}^{ab} = 2 g^2 \left[3 \left(\partial^X W \partial_X W +  \frac 1{\gamma^2} \partial^x W \partial_x
W \right) \left(\delta_c^5 + \delta_d^5\right)  - W^2 \right] \delta_c^{[a}\delta_d^{b]}.\label{cDW1}
\end{equation}
Using
\begin{eqnarray}
\dot W \equiv \partial_r W& = &\sqrt{\frac 23} \partial_r \varphi^\Lambda D_\Lambda P^s Q^s \cr
                          & = & - 3 g\left(\partial^X W \partial_X W +
                           \frac 1{\gamma^2} \partial^x W \partial_x W \right) \gamma,
\end{eqnarray}
(\ref{cDW1}) can be cast as
\begin{equation}
{\Omega_{cd}}^{ab} = 2 g^2 \left[- \frac{\dot W}{g\gamma} \left(\delta_c^5 + \delta_d^5\right)  -W^2 \right]
\delta_c^{[a}\delta_d^{b]}.\label{cDW2}
\end{equation}
The above expression is sufficient to show that the four dimensional slice is a space of non positive constant
curvature. First we observe that for a warped metric of the form (\ref{curvedmetric}), the curvature can be
written as
\begin{eqnarray}
& &{\Omega_{cd}}^{\bar a \bar b} =  {{\bar\Omega_{cd}}}^{\bar a \bar b} -
2(\dot A)^2 \delta_c^{[\bar a}\delta_d^{\bar b]}\label{cans1}\\
& & {\Omega_{cd}}^{\bar a 5} = -2 (\ddot A + (\dot A)^2) \delta_c^{[\bar a}\delta_d^{5]} \label{cans2}
\end{eqnarray}
where ${\bar\Omega}^{\bar a \bar b} = 1/2 e^{-2A} {\bar\Omega_{\bar c\bar d}}^{\bar a \bar b} e^{\bar c} \wedge
e^{\bar d}$ is the intrinsic curvature associated to the metric $g$. The comparison between (\ref{cans1}) and
(\ref{cDW2}) implies that ${\bar\Omega}^{\bar a \bar b}$ is proportional to $e^{\bar a}\wedge e^{\bar b}$  via a
function of $r$ {\it only}, that can be reabsorbed in the warp-factor. In practice this means that $A$ can be
taken {\it such that}
\begin{equation}
\left( g^2 W^2 - (\dot A)^2\right) e^{2A} = \frac 1{L^2},\label{warpf}
\end{equation}
 where $\bar R = -\frac{12}{L^2}$ the constant scalar curvature of $g$. It remains to demonstrate that $L\in {\cal R}$,
 i.e. is the length of $AdS_4$ (that reduce to Minkowski for $L=0$). From the comparison between (\ref{cans2})
and (\ref{cDW2}) it follows
\begin{equation}
\ddot A +(\dot A)^2 = g^2 \left(\frac{\dot W}{g \gamma} + W^2\right).
\end{equation}
Using (\ref{warpf}) we conclude that $\dot A = g\gamma W$, hence $(g^2 W^2 -(\dot A)^2),L^2 \ge 0$, because
$0\le\gamma^2\le 1$.

Let us summarize what we have presented in this section. It has been shown that  the well known domain wall
solutions are the unique BPS solutions that can be written as in (\ref{curvedmetric}) with the scalars depending
only on $r$.

In other words we have displayed  that assigning (\ref{dotvarphi}) is sufficient to get a domain wall. In this
way we can establish a one-to-one correspondence between the projector (\ref{PDW}) and domain wall solutions.

In the next section we will study a supersymmetric deformation of these solutions. In order to do so we will
focus on the flat domain walls (DW), i.e. $\gamma^2 = 1$. We conclude by observing, for future reference, that
in this case the metric (\ref{curvedmetric}) can be conveniently expressed as a {\it conformally flat} metric
\begin{equation}
\rmd s^2= \beta^2(x^5) \eta_{\mu\nu} \rmd x^\mu\rmd x^\nu,\label{mconf}
\end{equation}
where $x^5$ is related to $r$ by the change of coordinate $\rmd r = \beta(x^5) \rmd x^5$, with $\beta(x^5) =
e^{A(r)}$. The BPS equations become
\begin{eqnarray}
& &\frac{\dot\beta}{\beta^2} =  g \gamma W,\\
& &\dot \varphi^\Lambda = - 3g \beta \gamma \partial^\Lambda W,
\end{eqnarray}
the dot now indicating the derivative with respect to the new coordinate $x^5$.

\section{Null deformation}\label{null}

Now we want to consider together with (\ref{PDW}) the projector
\begin{equation}
\gamma_0 \epsilon_i = \pm \gamma_1 \epsilon_i.\label{PN}
\end{equation}
As will be shown clearly below, the resulting configuration can be seen as the generalization of the light-like
deformation of $AdS_5\times S^5$ studied in \cite{Chu:2006pa,Das:2006dz,Lin:2006ie}, from an effective five
dimensional point of view.\footnote{However, as for the non deformed DW, the uplifting to ten dimensions of our
5d model, is more involved than in the vacuum case, and is in general unknown.} For convenience, we name it as
``Null-deformed domain wall", or shortly NDDW, while we will refer to the  non deformed flat domain wall simply
as DW.

It is convenient to change our frame from the ordinary Minkowski to the lightcone one. We define $E^\pm \equiv
\frac{E^0 \pm E^1}{\sqrt 2}$, in order to have $\eta_{\pm\mp}= -1$. The (\ref{PN}) now reads $\gamma_\mp
\epsilon_i = 0$.

It is easy to verify that the two conditions over the covariant spinor are consistent. This point will be
discussed in section \ref{class} from the prospective of the classification method \cite{Gauntlett:2002nw}. We
will argue that, to some extent, the NDDW is the {\it most general non static  deformation} of the DW.

 The introduction of (\ref{PN}) reduces the amount of supersymmetry from $1/2$ to $1/4$ and, as a consequence the DW
metric (\ref{mconf}) is deformed. In order to study such deformation, we will consider the following metric (see
the appendix for more details)
\begin{equation}
\rmd s^2 = \beta^2(x^+, r)\left(-2k^2(x^+) \rmd x^+\rmd x^- + H(x^+,x^-,x^i,r) (\rmd x^+)^2 + \rmd r^2 + (\rmd
x^i)^2 \right).\label{METRIC}
\end{equation}
The above metric represents the most general light-like deformation of (\ref{mconf}), where the Minkowski slice
has been replaced by a generic PP wave and the conformal factor $\beta$ admits a dependence on lightcone
coordinate $x^+$. It reduces to the one studied in \cite{Chu:2006pa} for $\beta$ and $r$ taken to be
respectively the warp-factor and the radial coordinate of $AdS_5$ in the ``Brinkman form"
\cite[eq.(5)]{Chu:2006pa} respectively.

In order to attack the problem, we follow the same strategy as in the previous section. By first we discuss the
BPS equation for the scalars. Then, we use it to determine the curvature (to be compared with the one resulting
from the ansatz) via the integrability condition (\ref{GIC}).

First of all, we observe that, including both the gaugini and hyperini equations (\ref{gaugino}),
(\ref{hyperino}) a term of the form $\gamma^a \partial_a \varphi^\Lambda$, the projector  (\ref{PDW}) allows the
presence of a non zero $\partial_\pm \varphi^\Lambda$ component, which does not interfere with $\partial_5
\varphi^\Lambda$, remaining formally the same as for the DW. This means that equations  (\ref{prov}) and
(\ref{prow}) are untouched.

Similarly to (\ref{Dq}), we decompose $\partial_\pm q^X$ in (again the e.o.m imposes $K^X\partial_\mu q^Y
g_{XY}= 0$)
\begin{equation}
\partial_\pm q^X = v_s D^X P^s + u^X,
\end{equation}
 where $u^X$ is orthogonal to $K^X$ and $D^X P^s$. This decomposition is not only convenient for practical
 reasons, but also the two terms play different roles in the BPS equations (cfr. (\ref{pbeta})). This reflects their different origin:
 while $v_s D^X P^s$ is associated to the gauging, $u^X$ is related to the ungauged theory.

Taking into account that we want to study $1/4$-BPS configurations, it must be $v^r // \alpha^r$. Indeed
introducing another $\SU(2)$ direction is equivalent to add an extra projector condition like (\ref{PDW}), as
can be seen from the gravitini integrability condition (GIC) (\ref{GIC}).

The analysis of the gaugini equations goes along the same lines.

We can write the kinetic term of the scalars as
\begin{eqnarray}
& & \partial_a q^X = \left(2 \sqrt{\frac 32} g v \alpha_s D^X P^s + u^X \right) \delta_a^\pm + 2 \sqrt{\frac 32}
g \alpha_s D^X P^s \delta_a^5,\cr & & \partial_a \phi^x = \left(2 \sqrt{\frac 32} g w \alpha_s \partial^x P^s +
u^x \right) \delta_a^\pm + 2 \sqrt{\frac 32} g \alpha_s \partial^x P^s \delta_a^5. \label{Dvarphi}
\end{eqnarray}
As in hypermultiplet  case, the vector $u^x$ is orthogonal to $\partial^x P^s$ (similar considerations hold),
while the normalization of $v$ and $w$ is chosen for convenience to have $\partial_\pm q^X = v
\partial_5 q^X + u^X$ and $\partial_\pm \phi^x = w \partial_5 \phi^x + u^x$ respectively.

At first sight, the ``democratic" behavior of the scalars appearing in the DW case, eq.  (\ref{dotvarphi}),
(which is related to the success of the Fake supergravity approach) seems to be spoiled, because a priori $v$
and $w$ can be generic (unrelated) functions of the moduli space.

Following the same procedure as in the previous section, we can specialize the integrability condition
(\ref{GIC}) to the NDDW configuration, computing $D_aP^r$ and making use of (\ref{PDW}). The exact expression is
not so illuminating  and is presented in the appendix, (\ref{intgranddw}).

What is instead crucial, is that now the curvature $\Omega^{ab}$ (as well as the Ricci tensor) acquires
``off-diagonal" terms (i.e. not proportional to ${\delta_{[c}}^a{\delta_{d]}}^b$) related to the deformation.
Again this is a consequence of the new projection condition (\ref{PN}). The detail of this computation may be
found in the Appendix, equations (\ref{Opmpm}-\ref{Oab}).

Let us remark that the curvature is  completely determined by the integrality condition up to the
${\Omega_{\pm\tilde b}}^{\mp\tilde a}$ component. This feature is common to all the BPS solutions associated to
the projector (\ref{PN}). Indeed the component ${\Omega_{\pm\tilde b}}^{\mp\tilde a}$ always  cancels out
because it enters the integrability conditions multiplied by $\gamma_\mp$, that is zero on
$\epsilon_i$.\footnote{This does not happen for the other component of $\Omega^{\mp a}$ because of the symmetry
of the curvature. Furthermore we remind that, due to (\ref{PN}), $\gamma_{\pm\mp}\equiv
1/2[\gamma_\pm,\gamma_\mp]$ (that is not equal to $\gamma_\pm\gamma_\mp$) is not zero on $\epsilon$ but
proportional to the identity.}

Comparing the result we get from the GIC with the curvature computed starting by the ansatz (\ref{METRIC}), we
obtain the BPS equations:
\begin{eqnarray}
& & \frac{\dot\beta}{\beta^2} = g\gamma W,\ \ \ \ \ \ \ \ \ \ \gamma^2\equiv (-\alpha^s Q^s)^2 = 1,\label{dbeta}\\
& & \frac 1{\beta} (\frac{\dot\beta}{\beta^2})^\prime = -3 g^2 \left(v\, \partial^X W \partial_X W +
w\, \partial^x W \partial_x W \right), \label{pbeta}\\
& & \partial_i\partial_- H = \partial_-^2 H = \partial_- \dot H = 0. \label{H}
\end{eqnarray}
In force of eq.(\ref{H}) $H$ may be decomposed as:
\begin{equation}
H(x^+,x^-,x^i,r) = \tilde H(x^+,x^i,r) + H_-(x^+)\, x^-. \label{Hform}
\end{equation}
%The above equations will be discussed and their consequences studied in section \ref{anal}.
Let us note that the relation between the derivative with respect to $r$ and the superpotential, (\ref{dbeta}),
stays the same as in the DW case. In addition, we find again that $\gamma^2=1$. This is not surprising, in fact
as this is our input (as announced at the beginning, we restrict ourselves to null deformation of the flat
domain wall metric (\ref{mconf})) rather than a requirement of supersymmetry. Indeed, generalizing the metric
ansatz (\ref{METRIC}), it is possible to study curved domain wall deformation without changing the integrability
conditions (\ref{Opmpm}-\ref{Oab}). An other interesting remark relates to the absence of $u^\Lambda$ in the BPS
equations (\ref{dbeta}-\ref{H}). This is a first indication of the intrinsics difference between $u$ and $v$,
$w$-deformations.

However, the relation between $u^\Lambda$ and the metric comes from the Einstein equation ($(\pm\pm)\equiv (01)$
component, to be precise). As per usual, and as explained above, the first-order equations of light-like BPS
solution \cite{Gauntlett:2002nw},\cite{Celi:2003qk} are not sufficient to solve all the equations of motion and
fix the ansatz completely. Explicitly we find

\begin{eqnarray}
{R_\pm}^\mp & = & -9 g^2 \left(v\, \partial^X W \partial_X W + w\, \partial^x W \partial_x W \right) - u^\Lambda
u_\Lambda \cr &=& \frac 3{\beta} \left( D' - D \frac{2kk' + 1/2 \partial_- H}{k^2} + \frac 12 g\gamma W \dot H
\right) + \frac 1{2\beta^2} \left(\sum_i\partial_i^2 H + \ddot H \right),\label{Rpm}
\end{eqnarray}
where in parallel with (\ref{dbeta}) we introduce $D\equiv \frac{\beta'}{\beta^2}$. This equation is crucial to
relate the function $H$, characterizing the metric, to the scalars and the warp-factor, determining the
solution.

This is the only extra requirement coming from the equations of motion, (apart from $K^X\partial_a q^Y
g_{XY}=0$, used since the beginning) that otherwise are identically satisfied. Indeed, it is easy to verify that
the equations of motion for the scalars reduce to the one for the undeformed configuration, and, as in that
case, are identically satisfied. This result is somewhat expected because the null contribution to the kinetic
term is {\it traceless} thus does not enter in the laplacian (for the details of the calculation we refer the
reader to the appendix \ref{eom}).

The last non trivial constraint comes from the integrability conditions for the scalars (SIC). Taking a unifying
notation
\begin{equation}
\partial_\mu \varphi^\Lambda = \beta \left[ \left( -3g\gamma v_{(\Lambda)} \partial^\Lambda W + u^\Lambda
\right)\delta_\mu^+ -3g\gamma \partial^\Lambda W \delta_\mu^r \right],\label{SIC}
\end{equation}
where $$ v_{(\Lambda)} \equiv           \begin{cases} v, & \Lambda = 1,\dots,4n_H = X\\
                                        w, &\Lambda = 4n_H+1,\dots,4n_H+n_V = x+ 4n_H
                          \end{cases},$$
 the integrability condition $(\partial_+ \partial_r - \partial_r
\partial_-)\varphi^\Lambda = 0$ implies
\begin{multline}
\left(- 3g^2 W v_{(\Lambda)} + 9 g^2 \partial^\Sigma W\partial_\Sigma v_{(\Lambda)} +3 g\gamma D\right)
\partial^\Lambda W+ g\gamma W u^\Lambda =\\
9 g^2 \left(v_{(\Sigma)} - v_{(\Lambda)}\right) \partial^\Sigma W \partial_\Sigma \partial^\Lambda W +
3g\gamma\left(\partial^\Sigma W \partial_\Sigma u^\Lambda - u^\Sigma \partial_\Sigma \partial^\Lambda W\right),
\label{intsca}
\end{multline}
This expression will be discussed in section \ref{anal}, and will be explicitly solved for the simple models
studied in section \ref{sol}.

\subsection{Domain wall and classification}\label{class}

In this section we discuss a point that is in some sense {\it tangential} to the main stream of the paper. We
would like to shed some light on the relation between the DW solutions (and their deformations) and the
classification methods developed in \cite{Gauntlett:2002nw}, and successfully applied to supergravity theories
with 8 supercharges in
\cite{Gauntlett:2003fk,Gutowski:2003rg,Chamseddine:2003yy,Caldarelli:2003pb,Gutowski:2004ez,Gutowski:2004yv,Cariglia:2004kk,Gauntlett:2004qy,Cariglia:2004kk,Gutowski:2005id,Meessen:2006tu,Hubscher:2006mr,Jong:2006za,Bellorin:2006yr}.
In particular we want to understand within the framework of the classification, in which class the solutions we
are studying  fall in.\footnote{While this paper was being written, \cite{Gutowski:2006hk} and
\cite{Caldarelli:2006ww} appear. It contains some overlap with the discussion in this section.} Although some
facts and observations we report apply to diverse dimensions, we focus our discussion (as in the rest of the
work) on the 5d supergravity.

Let us emphasize  however that DW solutions in 5d gauged supergravity are only partially cover by the
classification method. Indeed no classification in gauged supergravity with hypermultiplet couplings currently
exists.

Moreover is intrinsically difficult to identify the DW  and all the solutions coming from the gauging, i.e. that
exist only in gauged supergravity (in the ungauged limit, $g\rightarrow 0$ reduces the vacuum). This occurs
because the classification method is essentially based on the ungauged theory. Indeed, the starting point of any
classification is to assume the existence of a covariantly constant spinor $\epsilon$. This can be divided into
two classes that are time-like or light-like. Such division implies the adoption of BPS solutions of ungauged
supergravity as a preferred base. To see this let us recall that a solution is said time-like or light-like if
the Killing vector $V^\mu$ constructed by the covariantly constant spinor $\epsilon$,
\begin{equation}
V^\mu \equiv 1/2\bar \epsilon^i\gamma^\mu\epsilon_i,\label{Vmu}
\end{equation}
 enjoys the former or the latter
properties, respectively. Following \cite{Gauntlett:2002nw}, the modulus of $V^\mu$ can be related via Fierz
identities to the scalar quantities $f\equiv 2 i \bar\epsilon^i\epsilon_i$. A crucial consequence of $\epsilon$
 being covariantly constant is that $V^\mu$ turns out to be Killing. Together with the Fierz identity
\begin{equation}
V^a\gamma_a \epsilon_i =  i f \epsilon_i,
\end{equation}
it implies the existence of preferred frame, in which a projector is associated to each BPS solution:
\begin{eqnarray}
& & \gamma_0 \epsilon = i \epsilon,\ \text{   if the (spinor, and by extension the) solution is time-like;}
\label{ptl}\\
& & \gamma_- \epsilon = 0,\ \text{  if the solution is light-like.} \label{pll}
\end{eqnarray}
The remarkable result \cite{Gauntlett:2002nw} is that these are the only projectors possible in the ungauged
supergravity\footnote{The full supersymmetry preserving solutions fit in the above classification but are
characterized by the existence of another covariantly constant spinor $\eta$ satisfying the complementary
projector, respectively $\gamma_0 \eta = - \eta$ and $\gamma_\pm \eta = 0$. Moreover, these are the unique
configurations belonging to the both classes.} (as a consequence the BPS solutions are or one half or maximally
supersymmetric). In this sense the classification method labels the configurations  by their origins in the
ungauged theory. This obviously is not all the story: the different solutions in the two classes are identified
by the allowed {\it Base spaces}.

It worth stressing that the projector (\ref{PDW}) associated to the domain wall can never be reduced to
(\ref{ptl}) or (\ref{pll}). Indeed, in the ungauged theory limit $g\rightarrow 0$ the algebraic condition
(\ref{PDW}) disappears and the domain wall reduces to maximally supersymmetric Minkowski vacuum. In the
classification contest the additional projector arises checking (the assumption of) the existence of covariantly
constant spinor. Indeed in the minimal gauged supergravity \cite{Gauntlett:2003fk} and in the gauged
supergravity with vector multiplets coupled \cite{Gutowski:2005id} the solutions are generically 1/4 BPS. From
this perspective, the BPS solutions of gauged supergravity are seen as deformations of the BPS configurations of
the ungauged gravity. Such a deformation is the result of the partial supersymmetry breaking introduced by the
gauging.

However, this point of view makes it difficult to characterized solutions like domain walls, which are
interesting in its own and, as we remarked, are exclusively a product of the gauging. It should be noted that
domain walls were not recognized in the classification up to now.

This gap can easily be filled by using the ``identification" between the DW and the projector (\ref{PDW}).
Indeed (\ref{PDW}) is only compatible  with the null projection (\ref{pll}) obtained in section \ref{DW}.
Assuming instead (\ref{ptl}), the anti-commuting algebra of $\gamma$-matrices is not realized on $\epsilon$. For
the same reason a projector of the form $\gamma_1\epsilon_i  = \theta^r {(\sigma_r)_i}^j \epsilon_j$
($\theta^r\theta_r =1$) is not compatible  with either (\ref{ptl}) or (\ref{pll}). This means that the
coordinate transversal to the wall can not be ``mixed" with time.

From these simple observations we learn that the DW can only belong to the class of light-like BPS solutions. At
the same time this implies something stronger: given a domain wall solutions the only supersymmetry preserving
(uncharged){\it deformations admitted are null} (the ones we consider in this work) or/and have their origin in
the coupling with hypermultiplets.

This statement reflects a peculiar point of view with respect the classification, in which the contribution of
the ungauged theory are seen as perturbation of the gauged solution. This is more useful when we are interested
in the properties of the latter.

Let us conclude by observing that the question over the existence of other deformations than the ones studied in
this paper seems to be strictly related to the existence of 1/8 BPS solutions.

\subsection{Analyzing the deformation} \label{anal}

In section \ref{null} we derived the BPS equation characterizing the NDDW. These equations will now  be analyzed
in order to understand the ``physics" behind them and construct explicit solutions (see section \ref{sol}). We
began by reminding the reader that a NDDW can be interpret in two ways. Indeed, as the name indicates can be
seen as a supersymmetric null deformation of a DW or as gauging deformations of an uncharged half BPS plane wave
configuration of the ungauged supergravity theory. These ``mother"  classes of solutions can be  derived
considering the projector (\ref{PDW}) and (\ref{PN}) separately. Their BPS equations are obtained from the
generic case by taking the limit $u^\Lambda,v_{(\Lambda)} \rightarrow 0$ and $g \rightarrow 0$,
respectively.\footnote{It is worth to note that no (plane wave) solutions associated to (\ref{PN}) and supported
by a non trivial potential ($W\neq$ constant) exist. This can be easily check by computing the equation of
motion for the scalar (\ref{emsca}) for the $x^+$ direction.} In the latter case, the kinetic term of the
scalars is simple given by $u^\Lambda$.

This fact points out the ``physical" difference between the null deformation controlled by $u^\Lambda$ and
$v_{(\Lambda)}$. We will refer to these as $u$-deformations and $v$-deformations, respectively.

The $u$-deformations are ungauged deformations, in the sense that $u^\Lambda$ identifies the scalar profile and
(up to some freedom in the function $H$, see section \ref{sol}) the metric of the (plane wave) solution in
$g\rightarrow 0$ limit. The $v$-deformations are instead a product of the gauging, and are the unique ones
related to the potential. It follows from eq.(\ref{pbeta}), that the potential can be {\it time-dependent} (via
$x^+$) only in presence of $v$-deformations. This makes it very appealing to construct these kind of solutions.

However, it seems very hard to obtain explicit solutions in the most general set-up of section \ref{null}, at
least analytically. The major difficulty to overcome is the integrability condition of the scalars
(\ref{intsca}).

Indeed the recipe for constructing a solution consists of:
\begin{enumerate}
\item assigning the matter sector and the {\it gauging}, in practice giving a prepotential $W$; \item obtaining
from the SIC (\ref{intsca}) admissible $v_{(\Lambda)}$ and $u^\Lambda$ as function of the moduli; \item
integrating the scalar BPS equations (\ref{SIC}) and determining $\beta$, $v_{(\Lambda)}$  and  $u^\Lambda$ as
functions of spacetime ($x^+$ and $r$); \item deriving $H$ from (\ref{Rpm}).
\end{enumerate}

For the first step, we note that the orthogonality between $u^\Lambda$, the Killing vector $K^X$ and the
$\SU(2)$-covariant derivative of prepotential $D^\Lambda P^s$ requires  (for $u^\Lambda \neq 0$) $n_V,n_H \neq
1$.

The second step is certainly the most delicate. The SIC can be interpreted as an implicit definition of
$v_{(\Lambda)}$ and $u^\Lambda$ that otherwise do not have an well-known geometric origin as $W$.

The difficulties of points 3 and 4 are of technical nature, and at worst can be faced using numerical methods.

The plan above will be applied in the next section, considering $u$-deformation only. In this special case, more
can be said on the solution. First of all, (\ref{pbeta}) means that $W=\frac{\gamma}{g} \frac{\dot
\beta}{\beta^2}$, $\gamma=\pm 1$, is a function of $r$ only. Moreover (\ref{pbeta}) tells us that
$\frac{\beta^\prime}{\beta^2}= D(x^+)$ or, in other words, that the warp-factor decomposes as follows
\begin{equation}
\beta^{-1} = f(r) + g(x^+).
\end{equation}

Before concluding let us add a comment on the SIC (\ref{intsca}). Due to the orthogonality between $u^\Lambda$
and $\partial^\Lambda W$ it actually corresponds to two distinct equations. That in the $u^\Lambda$ direction
can be interpret as the definition of $u^\Lambda$ (and $v_{(\Lambda)}$), or in other words is the consistency
condition between the gauging and the ungauged solutions. That in the $\partial^\Lambda W$ direction determines
$D$, i.e. the dependence on {\it time} ($x^+$) of the warp-factor $\beta$. Taking a constructive point of view,
the first equation determines whether, for each of the possible direction orthogonal to $\partial^\Lambda W$, is
possible to adjust the modulus of $u^\Lambda$ in order to find a solution. In the examples we present in section
\ref{sol} this occurs. Furthermore it turns out that $u^\Lambda u_\Lambda$ is not completely determined by the
SIC.

\section{Explicit solutions}\label{sol}
In this section we present the explicit realization of a NDDW for the simplest models we can consider. For this
purpose we restrict ourselves to $u$-deformation.

 Indeed, as discussed above, we need {\it at least} $n_V\ge2$ or/and $n_H\ge2$. For example it is not possible to realize
  the  orthogonality of $\partial^\Lambda W$ and $u^\Lambda$ in a trivial way, i.e. taking one lying in the
  Hypergeometry and other in the Very Special geometry. Indeed, the integrability condition of the scalars
  (\ref{intsca}), would force the solution to reduces to the plane way of the ungauged theory or, alternatively to a flat
  domain wall.\footnote{This result confirms, in accordance with the expectation, that the ungauged matter sector (the one
   where $\partial W = 0$) decouples from the gauged matter sector and do not contribute to the solution. In other words given
   a domain wall supported by $n_V$ vector multiplets and $n_H$ hypermultiplets, an arbitrary number of constant
   matter multiplets  can always be added.} In that follows,
  we will focus over the cases: a) $(n_V,n_H)=(0,2)$; b)  $(n_V,n_H)=(2,0)$. In particular we consider the group
  manifolds $\frac{\Sp(2,1)}{\Sp(2)\times \Sp(1)}$ and $\frac{\SO(2,1)}{\SO(2)}$.

The solutions we obtain are peculiar because the warp-factor $\beta$ turns out to be a function of $r$ only,
remaining untouched by the deformation. This feature depends only on the special gauging chosen in order to
guarantee the existence of analytic solutions. Why this happens is clarified in the appendix \ref{adapt} by
means of the adopted coordinates \cite{Celi:2004st}.

\subsection{A $n_H=2$ solution: the $\frac{\Sp(2,1)}{\Sp(2)\times \Sp(1)}$ model.}\label{solh}

%XXX INTRODUZIONE AL MODELLO: scegliere quali dettagli mettere qui e quali in appendice XXX
Details on the geometry and coset parametrization of the coset space $\frac{\Sp(2,1)}{\Sp(2)\times \Sp(1)}$ are
given in the appendix \ref{app:paramCosets}. The space is characterized by the following metric
\begin{equation}
\rmd s^2= (\rmd h)^2+ (B^1)^2+(B^2)^2+(B^3)^2 +2\rme^{-h}\left[{ (\rmd e^0)^2+(\rmd e^1)^2+(\rmd e^2)^2+(\rmd
e^3)^2 }\right]. \label{sp_metric}
\end{equation}

In order to get a simple configuration, we consider as isometry to be gauged a translation. Because the metric
 (\ref{sp_metric}) is cyclic in the $b^r$, we take
\begin{equation}
K=\partial_{b^1}.
\end{equation}

In order to compute the prepotential $P^r$, we follow the same strategy as in \cite{Achucarro:2005vz}. Indeed,
for practical purposes, is convenient to use another definition of $P^r$ different than (\ref{KillingP}). A
Killing vector preserves the connection $\omega^r$ and K{\"a}hler two forms $J^r$ ($\frac 12 \nu J^r\equiv R^r$,
with $\nu=-1$ in our paper) only modulo an $\rSU(2)$ rotation. Denoting by $\cL_\Lambda $ a Lie derivative with
respect to $k_\Lambda $, we have
\begin{equation}
\cL_\Lambda \omega^s  = -\ft12 \nabla r^s_\Lambda ,\quad \cL_\Lambda J^r=\varepsilon^{rst}r^s_\Lambda J^t,
\label{Liederinr}
\end{equation}
where $r^s_\Lambda$ is known as an $\rSU(2)$ {\em compensator}. The $\rSU(2)$-bundle of a quaternionic manifold
is non-trivial and therefore it is  impossible to get rid of the compensator $r_\Lambda^s$ by a redefinition of
the $\rSU(2)$ connections.\footnote{This is in contrast with $\cN=2$ rigid supersymmetry, since hyper-K{\"a}hler
manifolds have a trivial $\rSU(2)$ bundle, and therefore no compensator.} The moment map can be expressed in
terms of the triplet of connections $\omega^s$ and the compensator $r^s_\Lambda$ in the following way
\cite{Galicki:1987ja}:
\begin{equation}
\cP^s_\Lambda =\ft12 r^s_\Lambda+\iota_\Lambda \omega^s. \label{momentmap}
\end{equation}

 For this Killing vector the compensator turns up to be zero, and the moment map $P^r$, following (\ref{momentmap}), is
\begin{equation}
P^r=\iota_K w^r= - \frac 12 e^{-h}\delta^{1r}.
\end{equation}
Accordingly with the BPS condition, we can choose as $u$ any vector field in $\frac{\Sp(2,1)}{\Sp(2)\times
\Sp(1)}$ orthogonal to $K$ and $\iota_K J^r$, for example:
\begin{equation}
u = f\left(- \partial_{e^0} + e^r \partial_{b^r}\right). \label{u_H}
\end{equation}
The $f$ at this stage is arbitrary function of the scalar manifold, but the integrability condition of the
scalars will fix its dependence on the ``running" ones (the others are irrelevant for determining the final
solution). Considering only $u$-deformations, the SIC (\ref{intsca}) becomes
\begin{equation}
3 D \partial^X W + W u^X = 3 \left(\partial^Y W \partial_Y u^X -u^Y \partial_Y \partial^X W \right).
\end{equation}
As the superpotential $W$ is a function of the Cartan coordinate $h$ (with respect to  which the metric
(\ref{sp_metric}) is  by definition diagonal)  {\it only}, the above equation implies
\begin{eqnarray}
& & D = 0 = \beta' \Longrightarrow \beta = \beta(r), \cr & & \partial_h \ln f = -\frac 13 \Longrightarrow f =
{\cal C} F(q) e^{-h/3}, \ \ \ \ \  \partial_h F(q) = 0.
\end{eqnarray}
 The constant ${\cal C}$ has been introduced for convenience. Without loosing of generality we can restrict
 $F(q)$ to be a function of $e^0$ and $b^r$ only. As we will see, $F$ takes the role of {\it generating
 function} of the solution. Indeed, eq.(\ref{SIC}) gives
 \begin{eqnarray}
 & & {e^0}^\prime = - \beta f = -F, \\
 & & {b^r}^\prime = \beta f =  F.
 \end{eqnarray}
 The above equations imply $$ b^r= -e^r e^0 + C^r, \ \ \ \ \ \  C^r= \text{constant}, $$ and
 \begin{equation}
 \rmd x^+ = - \frac{\rmd e^0}{F(e^0, -e^r e^0 + C^r)},
 \end{equation}
which can always be integrated and inverted {\it piecewise} for a smooth $F(e^0,b^r)$.\footnote{However,  for a
generic $F$ the solution will develop a singularity of similar kind as in \cite{Chu:2006pa,Das:2006dz}.}

Let us remark that the feature $\partial_+ \beta = 0$ is not generic but a consequence of the simple model we
have chosen. This property allows us to integrate immediately eq.(\ref{dbeta}):
\begin{eqnarray}
& & h=\frac 32 \ln\left[\frac{2g\gamma}{\sqrt 6 {\cal C}} (r-r_0)\right], \\
& & \beta= \frac 1{\cal C} e^{h/3} = \frac 1{\cal C} \left[\frac{2g\gamma}{\sqrt 6 {\cal C}}
(r-r_0)\right]^{1/2}.
\end{eqnarray}
Before discussing the $x^+$-dependence of the solution, let us remark that at $r=r_0$ the solution has a
singularity. We may take $r_0=0$ and chose $\gamma$ in order to have the solution defined for $r>0$. The
singularity exists even when the deformation is absent. Indeed the superpotential
$$
W = \frac 1 {\sqrt 6}\left[\frac{2g\gamma}{\sqrt 6 {\cal C}} r \right]^{-3/2},
$$
that is related to the curvature by the BPS equations, explodes for $r = 0$. This is not surprising because this
happens for all DWs obtained by the gauging of a translation and for this specific model the radial dependence
is unaffected by the deformation.\footnote{For this specific gauging more can be said about the stringy origin
of the solution. In Calabi-Yau compactification the Cartan modulus is associated to the Volume $V$ of the
compact space (to be precise $V \propto e^h$)\cite{Ceresole:2001wi}. In this specific case the singularity
occurs when the CY shrinks to zero and the supergravity approximation is breaking down.}

As the warp-factor is independent of the deformation, $u$ and the dependence on $x^+$ enter the metric only
through the function $H$, describing the ``wave". As explained in the previous sections, $H$ is determined by
(\ref{Rpm}) using the decomposition in (\ref{Hform}):
\begin{equation}
\frac 32 \frac Xr + \dot X + Y = B r^{-3/2},
\end{equation}
where $X \equiv \dot {\tilde H}$, $Y\equiv \sum_i \partial_i^2\tilde H$ and $B \equiv - 4
\left[\frac{2g\gamma}{\sqrt 6 {\cal C}}\right]^{-3/2} F^2$, $B<0$. This equation can be easily integrated in $r$
under the assumption (not required by supersymmetry) $\partial_i \dot {\tilde H} = 0$, that implies $Y=Y(x^+)$.
We get
\begin{equation}
H= -2 \rho r^{- 1/2} + 2 B r^{1/2} - \frac 15 Y r^2 + \sigma + H_- x^-,
\end{equation}
with $\rho$ and $H_-$ generic functions of $x^+$, $\rho=\rho(x^+)$, $H_-=H_-(x^+)$, and $\sigma=\sigma(x^+,x^i)$
such that $\sum_i \partial_i^2 \sigma = Y$.

Let us observe that $H$, measuring the light-like deformation of the metric, is only partially controlled by the
shape of $F$ (via $B$), measuring the deformation of the scalar sector. Indeed, the presence of less
supersymmetry preservation with respect to the DW allows more freedom and, in contrast with the DW case,
different metrics can correspond to a single scalar profile.

Taking $F= e^0$ and fixing all the integration constant (and functions) to a convenient value, a simple solution
is
\begin{eqnarray}
& & \rmd s^2 = \frac 23 g r\left(-2 \rmd x^+\rmd x^-  - 8 g \sqrt r e^{-2x^+} (\rmd x^+)^2 + \rmd r^2 + (\rmd x^i)^2 \right),\\
& & h =\frac 32 \ln[g r],\\
& & e^0 = e^{-x^+},
\end{eqnarray}
with the other scalars identically zero. This solution exhibits a light-like singularity for $x^+ \rightarrow -
\infty$.

\subsection{A $n_V=2$ solution: the $\frac{\SO(2,1)}{\SO(2)}$ model.}

We now consider the moduli space  ${\cal M} = {SO(1,n_V) \over SO(n_V)}, n_V >1.$ We will use the
parametrization in \cite{Cosemans:2005sj}. We can then take the following polynomial
\begin{equation}
N(h)={3\over
2}\sqrt{{3\over2}}\left(\sqrt{2}h^{0}(h^{1})^2-h^{1}\left[(h^{2})^2+\ldots+(h^{n_V})^2\right]\right)\,
.\end{equation}
 This means that the non-vanishing components of the tensor $C_{IJK}$ are
 \begin{equation}
C_{011}={\sqrt{3}\over 2}\, ,\quad C_{1ab}=-{\sqrt{6}\over 4}\, \delta_{ab}\, , \,a,b = 2,\ldots,n_V\, .
\end{equation}
 The constraint $N=1$ can be solved by
 \begin{eqnarray} h^{0}&=&\sqrt{{2\over
3}}\left(\frac{1}{\sqrt{2}(\varphi^1)^2}+\frac{1}{\sqrt{2}}\varphi^1
\left[(\varphi^2)^2+\ldots+(\varphi^{n_V})^2\right]\right),\\
h^{1}&=&\sqrt{\frac{2}{3}}\varphi^1, \qquad h^{a}=\sqrt{\frac{2}{3}}\varphi^1 \varphi^a.\end{eqnarray}

Applying the Very Special geometry identities (\ref{identVS}), the metric $g_{xy}$ results diagonal in this
parametrization,
\begin{equation}
g_{xy} = Diag(\frac 1 {(\phi^1)^2}, \frac{(\phi^1)^3}3,\ldots,\frac{(\phi^1)^3}3).
\end{equation}
Here we are interested in performing a $\U(1)$ gauging. The constraint (\ref{constraint}) implies for the
constant $P_I^r$, ${\vec P_I \times \vec P_J=0} $, therefore the prepotential is $P^r = P^r_Ih^I$ with $P_I^r =
V_I Q^r$. It follows $W= \sqrt{\frac 23} V_I h^I$. In order to get an analytic solution we choose $V_I=
V\delta_I^1$. Explicitly
\begin{equation}
W= \frac 23 V \phi^1. \label{Wnv}
\end{equation}
According to the orthogonality condition we can take
\begin{equation}
u^x = f \delta_2^x.
\end{equation}
Due to the $\SO(n_V)$ symmetry of the moduli space, $u^x$ may always be cast in  this form. This means that, for
the special gauging (\ref{Wnv}) we can restrict without loss of generality to $n_V=2$. As in the previous
section we start by analyzing the integrability conditions for the scalars. The equation in the direction $1$
gives
\begin{equation}
3D\partial^1 W= - 3 f \partial_2\partial^1 W = 0, \ \ \ \ \Rightarrow \ \ \ \ D=0,\label{1nv}
\end{equation}
while the equation along the second component determines $f$
\begin{equation}
W f = 3 \partial^1 W\partial_1 f, \ \ \ \ \Rightarrow \ \ \ \  \partial_1 \ln f = \frac 1{3\phi^1}.\label{2nv}
\end{equation}
Again (\ref{1nv}) entails $\beta=\beta(\phi^1)=\beta(r)$ while from (\ref{2nv}) follows $$f= F(\phi^2)
\beta^{-1}= {(\cal C)}^{-1} F(\phi^2) {(\phi^1)}^{1/3}.$$ As in the hypermultiplet example $F$ is completely
arbitrary.

The profile of $\phi^1(r)$ can be easily determined integrating (\ref{dbeta}):
\begin{equation}
\phi^1 = \left[\frac 43 g \gamma {\cal C} V r\right]^{-3/2}, \ \ \ \ \beta= {\cal C}\left[\frac 43 g \gamma
{\cal C} V r\right]^{1/2}.
\end{equation}
Accidentally the solution turns out to be practically identical to one obtained in the previous section.

\section{Discussion}\label{conclu}

In this paper we analyzed null deformations of flat domain wall solutions (NDDW) in gauged supergravity. In our
study we used an approach mainly based on the choice of an ansatz explicitly showing, however, that we covered
all the solutions of the class we were interested in. In this respect, we reviewed and further investigated the
relation between the projector (\ref{PDW}) and flat domain wall solutions (DW) (for related discussions see e.g.
\cite{Ceresole:2001wi,Chamseddine:2001hx,LopesCardoso:2002ec}). This allowed us to identify the DW solutions as
light-like in the classification framework, and, more important, it allowed us to prove that {\it all the
possible deformations of the DW have origin in the hypermultiplet sector} or/and {\it are null}.

We showed that the null deformations can have a ``gauging" ($v$-deformation) or a ``non gauging"
($u$-deformation) nature. This conceptual difference has practical consequences: only the presence of a
$v$-deformation can give rise to a time-dependent (super)potential.

As the superpotential $W$ controls the dependence of the scalars (and, via backreaction, of the metric) on $r$,
$u^\Lambda$ and $v_{(\Lambda)}$ determine the lightcone time dependence. However, in comparison to $W$ they do
not have an intrinsic geometrical origin on the moduli space. $v_{(\Lambda)}$ and $u^\Lambda$ (or better
$u^\Lambda u_\Lambda$) acquire a well-defined meaning once they satisfy the integrability conditions of the
scalars (SIC) as spacetime functions, $\varphi^\Lambda=\varphi^\Lambda(r,x^+)$.

The SIC play a crucial role in constructing solutions. We showed how they can be solved, and two analytical
solutions supported by scalars in the hypermultiplet and in the vector sector respectively were found.

Our study also provided insights that seems to apply to generic BPS solutions in gauged supergravity
\cite{hypereq}. We note for the  first time that the compatibility of gauging imposes restrictions on the number
of matter multiplets, even at the level of an abelian gauged group. Indeed if we
 consider $u$-deformation, the resulting solution can be equivalently considered as the outcome of the soft
 supersymmetry breaking produced by the gauging on a (null) background of the ungauged theory (identified by $u^\Lambda$).
  The resulting  condition
 for preserving supersymmetry, $u^\Lambda D_\Lambda \vec P = u^X K_X = 0$, forces the number of matter multiplets to be
 different by one, $n_H,n_V\neq 1$.
While for vector multiplets (where there is one scalar in each multiplet) such condition is meeting the naive
expectation that for each ``active" spacetime direction there is {\it at least} one scalar flowing (an
expectation that can be made rigorous using the adapted coordinate of Appendix \ref{adapt}), this is far less
obvious in the hypermultiplet case (where there are four scalars in each multiplet) and completely unexpected
when both kinds are present. The above consideration reinforces the idea that it is more ``natural" to regard
the scalars of a hypermultiplet as a unique quaternionic scalar.

At the same time, this is an indication that an extension of the Fake supergravity formalism may be possible, at
least for $u$-deformed DW. Roughly speaking, one expects that, in analogy with the DW, the supergravity can be
effectively described by two scalars, encoding respectively $x^+$ and the $r$ dependence. The second scalar
should mimic only the scalars (the multiplets) involved in the gauging. Such a splitting should also appear in
the fake BPS conditions. In support of this picture, we find that the democratic treatment for the scalars,
introduced in \cite{Ceresole:2001wi} and extended for curved domain walls in \cite{Celi:2004st}, perfectly works
also for NDDW.

Our results raise interesting questions that have only been considered briefly.

First of all, it would be very appealing to explore the holographic meaning of NDDW. Assuming the validity of
gauged/gravity correspondence (at least when the gravity background is asymptotically AdS), one would expects
that, being the deformation of $AdS_5$ associated to a DW and null deformation compatible at the supergravity
level, the same should happen for the corresponding deformations of $\cN=4$ SYM. Would be very interesting to
check this explicitly at the gauge theory level.

The ultimate question that naturally arises is whether the flow in (the analogue of) the radial coordinate still
describes the RG flow in the dual field theory. In order to address this problem, one should take the ($\cN=2$
embedding of the) kinks that have dual known flow and construct their null deformation. Constructing such
solutions explicitly is certainly not an easy task.

More concretely, let us consider the FGPW flow. Its $\cN=2$ embedding has been given in \cite{Ceresole:2001wi}
in terms of one vector multiplet, one hypermultiplet with a gauging of a $\U(1)\times \U(1)$ symmetry of the
scalar manifold. However, as commonly happens in presence of compact gauging with both hypers and vectors
coupled, the actual solution (out of the fixed points) is known only numerically. This circumstance
unfortunately makes it very difficult to solve the SIC and to construct a consistent null deformation.

On other hand, it is relatively easy to construct NDDWs based on non compact gauging, as was shown in section
\ref{sol} and it should be even possible to obtain their uplifting. However, it is more difficult to find the
holographic dual of such configurations because they are deformations of DWs that are not asymptotically AdS.

A possible way of circumventing such difficulties could be achieved  developing a generalized Fake formalism, on
the lines discussed above. Furthermore, using the technique presented in \cite{Papadimitriou:2006dr}, one could
 obtained non supersymmetric but stable NDDW.

\smallskip

A related problem would be to consider null deformation of curved domain wall
\cite{LopesCardoso:2001rt,Chamseddine:2001hx,LopesCardoso:2002ec,Behrndt:2002ee,LopesCardoso:2002ff}. This
extension can be done by simply generalizing the metric ansatz (\ref{METRIC}), as our calculations are valid for
a generic $\gamma$, with $\gamma^2\leq 1$. This would make possible to consider null deformation of solutions
like ``Janus" \cite{Bak:2003jk}, which is conjectured to be dual to an interface CFT \cite{Clark:2004sb}. The
stability of this ten-dimensional Type $0$ solution was proven in \cite{Freedman:2003ax} using Fake
supergravity, while its embedding into $\cN=2$, $d=5$ gauged supergravity has been derived in
\cite{Clark:2005te}, following \cite{Celi:2004st}. The supersymmetric Type II Janus has been recently obtained
in \cite{D'Hoker:2006uu} and its holographic interpretation discussed in \cite{D'Hoker:2006uv}.

\smallskip
A completely different application of the NDDW would be in the study of possible supersymmetric decay of domain
walls. Very recently, in \cite{Skenderis:2006rr} it has been found that stable domain walls can asymptote to
unstable anti-de Sitter vacua. The authors conjectured that these solutions decay via a time-dependent process
to some near-by stable domain wall. It would be interesting to see whether a NDDW might represent a possible
decay channel.

Finally, another point that deserves further attention is the existence of 1/8 BPS deformations of DWs. The very
recent result of \cite{Bellorin:2006yr} suggests that solutions preserving only 1/8 of supersymmetry exist
already in ungauged supergravity with hypermultiplet couplings. In contrast, without hypermultiplet coupled the
supersymmetric configuration preserves {\it at least} two supercharges \cite{Gutowski:2005id}. It is then
reasonable to assume these deformations should exist. Characterizing such configurations is interesting on its
own right and could help in the arduous task of classifying all the BPS solutions of $\cN=2$ gauged supergravity
with hypermultiplets coupled.

\medskip
\section*{Acknowledgments.}

\noindent The author thanks Aldo Cotrone for the support during the preparation of this work. He is graceful to
Roberto Emparan, Joaquim Gomis, Diederik Roest, Jorge Russo and Geert Smet for the useful discussions.
Furthermore, he thanks Paul Smith for revising an improved version of this manuscript. He blesses Antonella for
the arrival of Martina, their marvellous daughter who arrived recently.

AC is supported by the ``Ministerio de Educaci\'on y Ciencia", Spain through the ``Programa Juan de la Cierva".
This work is supported in part by the European Community's Human Potential Programme under contract
MRTN-CT-2004-005104 `Constituents, fundamental forces and symmetries of the universe', by the Spanish grant MCYT
FPA 2004-04582-C02-01 and by the Catalan grant CIRIT GC 2005SGR-00564.

\newpage

\appendix

\section{Metric and Integrability conditions}\label{MIC}

Inspired by \cite{Chu:2006pa}, we choose the following metric ansatz\footnote{With respect to \cite{Chu:2006pa}
we define the lightcone coordinate differently, in order to have $\eta_{\pm\mp}= 1$. The main difference is that
here the warp-factor $\beta$ is a generic function of $r$ and $x^+$. The deformed AdS metric of
\cite{Chu:2006pa} is recovered for $\beta= 1/r$.} (in a conformal gauge):
\begin{eqnarray}
&& \rmd s^2 = \beta^2(x^+, r)\left(-2k^2(x^+) \rmd x^+\rmd x^- + H(x^+,x^-,x^i,r) (\rmd x^+)^2 + \rmd r^2 +
(\rmd x^i)^2 \right), \label{metric}\\
&& E^{\pm} =\beta \rmd x^+,  \ \ \ \ \  E^{\mp} = \beta (k^2\rmd x^- - 1/2 H \rmd x^+),\ \ \ \ \ E^i = \beta
\rmd x^i, \ \ \ \ \  E^5 = \beta \rmd r.
\end{eqnarray}
It follows
\begin{eqnarray}
& & {w^\pm}_\pm = \left(\frac{\beta^\prime }{\beta^2} + \frac{2kk^\prime  + 1/2 \partial_- H}{\beta k^2}\right)
E^\pm,
\ \ \ \ \  {w^\pm}_r = \frac{\dot \beta}{\beta^2 } E^\pm, \\
&& {w^\mp}_i = - 1/2 \frac{\partial_i H}{\beta} E^\pm + \frac{\beta^\prime }{\beta^2} E^i, \ \ \ \ \
{w^\mp}_r = -1/2 \frac{\dot H}{\beta} E^\pm + \frac{\dot \beta}{\beta^2 } E^\mp +  \frac{\beta^\prime}{\beta^2} E^5, \\
& & {w^i}_r = \frac{\dot \beta}{\beta^2 } E^i,
\end{eqnarray}
where we indicate the derivative with respect to the spacetime coordinates $x^+$ and $r$ with a prime and a dot,
respectively: $\beta^\prime \equiv \partial_{x^+} \beta$, $\dot \beta \equiv \partial_r \beta$. For the
curvature we have
\begin{eqnarray}
\Omega^{\pm\mp} &=& \frac{E^\pm}{\beta^2} \wedge\left[\left(\frac{\partial^2_- H}{2k^2} - (\dot
\beta/\beta)\right)^2 E^\mp + \frac{\partial_i\partial_- H}{2 k^2} E^i + \left(\frac{\dot\beta^\prime\beta - 2
\dot \beta \beta^\prime}{\beta^2}
 + \frac{\partial_- \dot H }{2 k^2}\right) E^5\right], \\
\Omega^{\pm i} &=& -(\dot\beta/\beta^2)^2 E^\pm\wedge E^i,\ \ \ \ \
\Omega^{\pm r} = - \frac{\ddot\beta \beta - 2 \dot\beta^2}{\beta^4} E^\pm\wedge E^5,  \\
\Omega^{\mp i} &=& 1/\beta^2 E^\pm\wedge\left\{\frac{\partial_-\partial_i H}{2 k^2} E^\mp +
 \left[\left(\frac{  \beta^{\prime\prime} \beta -2(\beta^\prime)^2}{\beta^2} - \beta^\prime/\beta \frac{2 k k^\prime +
 1/2 \partial_- H}{k^2} + \frac{\dot \beta \dot H}{2\beta}\right)\delta_j^i \right. \right.\cr &+& \left.
 \left.
 1/2 \partial_{ij} H \right] E^j
+  1/2 \partial_i \dot H E^5\right\} -\left(\frac{\dot\beta}{\beta^2}\right)^2 E^\mp\wedge E^i -
 \frac{\dot\beta^\prime - 2\beta^\prime \dot \beta}{\beta^4} E^i\wedge E^5,\\
\Omega^{\mp r} &=& 1/\beta^2 E^\pm \wedge \left[\left(\frac{\partial_- \dot H}{2 k^2} + \frac{\dot\beta^\prime -
2 \dot\beta\beta^\prime}{\beta^2}\right) E^\mp + 1/2 \partial_i \dot H E^i + \left(1/2 \ddot H +
\frac{\dot\beta\dot H}{2 \beta } + \frac{\beta^{\prime\prime} \beta - 2 (\beta^\prime)^2}{\beta^2}
\right.\right.\cr &-& \left.\left.\beta^\prime \frac{2 k k^\prime + 1/2 \partial_- H}{k^2}\right) E^5\right]
- \frac{\ddot \beta \beta - (\dot\beta)^2}{\beta^4} E^\mp \wedge E^5,\\
\Omega^{ir} & = & \frac{\dot\beta^\prime\beta - 2\dot\beta \beta^\prime}{\beta^4} E^\pm\wedge E^i - \frac{\ddot
\beta\beta- (\dot\beta)^2}{\beta^4} E^i\wedge E^5,\ \ \ \ \ \Omega^{ij} = -
\left(\frac{\dot\beta}{\beta^2}\right)^2 E^i\wedge E^j.
\end{eqnarray}
In that following we will use the notation $\tilde  a$ to indicate the flat indeces different than $0$, $1$ ($
\pm$, $\mp$).

The functions in the ansatz will be determined by the comparison with the gravitini integrality conditions (GIC)
and the equations of motion for the metric. The formers are established studying the consistency of (\ref{GIC})
with the projectors (\ref{PDW}) and (\ref{PN}). Explicitly, from (\ref{Dvarphi}) it follows
\begin{eqnarray}
 D_a P^r &=& \partial_a \varphi^\Lambda D_\Lambda P^r = \cr
&=& \sqrt{\frac{3}{2}} g \left[3\left(v\, \partial^X W \partial_X W + w\, \frac 1{\gamma^2} \partial^x W
\partial_x W \right) \delta_a^\pm + 3\left(\partial^X W \partial_X W +
 \frac 1{\gamma^2} \partial^x W \partial_x W \right)\delta_a^5\right] \alpha^r, \cr
\gamma &\equiv& - \alpha^s Q^s,
\end{eqnarray}
that implies
\begin{eqnarray}
\left\{1/2{\Omega_{cd}}^{ab}\right. &+&\left. 6g^2\left[v\, \partial^X W \partial_X W +
 w\, \frac 1{\gamma^2} \partial^x W \partial_x W \right] \delta_{[c}^\pm\gamma_{d]}\gamma_5 \right.\cr \left.\right.
&-& \left.g^2 \left[3\left(\partial^X W \partial_X W + \frac 1{\gamma^2} \partial^x W \partial_x W \right)
\left(\delta_c^5 + \delta_d^5\right) -
 W^2\right]\gamma_{cd}\right\}\epsilon_i = 0.   \    \label{intgranddw}
\end{eqnarray}
The above equation fixes the curvature to be
\begin{eqnarray}
\Omega^{\pm\mp} &=& - g^2 W^2  E^\pm \wedge E^\mp - 3 g^2 \left(v\, \partial^X W \partial_X W +
 w\, \frac 1{\gamma^2} \partial^x W \partial_x W \right) E^\pm\wedge E^5,\label{Opmpm}\\
\Omega^{\pm\tilde a} &=& g^2 \left[3 \left(\partial^X W \partial_X W +
 \frac 1{\gamma^2} \partial^x W \partial_x W \right) \delta_{\tilde a}^5 -
 W^2\right] E^\pm\wedge E^{\tilde a},\label{Opma}\\
\Omega^{\mp\tilde a} &=& {\Omega_{\pm\tilde b}}^{\mp\tilde a} E^\pm \wedge E^{\tilde b}  - 3 g^2 \left(v\,
\partial^X W \partial_X W +w\, \frac 1{\gamma^2} \partial^x W \partial_x W \right) \delta_5^{\tilde a}
E^\pm\wedge E^\mp \cr & +& g^2 \left[3 \left(\partial^X W \partial_X W + \frac 1{\gamma^2} \partial^x W
\partial_x W \right) \delta_{\tilde a}^5 - W^2\right] E^\mp\wedge E^{\tilde a} \cr
&+& 3 g^2 \left(v\, \partial^X W \partial_X W +
 w\, \frac 1{\gamma^2} \partial^x W \partial_x W \right) E^{\tilde a}\wedge E^5,\label{Ompa}\\
\Omega^{\tilde a\tilde b} &=& g^2 \left[3 \left(\partial^X W \partial_X W + \frac 1{\gamma^2} \partial^x W
\partial_x W \right) \left(\delta_{\tilde a}^5 + \delta_{\tilde b}\right) - W^2\right] E^{\tilde a}\wedge
E^{\tilde b} \cr &-& 6 g^2 \left(v\, \partial^X W \partial_X W + w\, \frac 1{\gamma^2} \partial^x W \partial_x W
\right) \delta_{\tilde c}^{[\tilde a} \delta_5^{\tilde b]} E^\pm\wedge E^{\tilde c}.\label{Oab}
\end{eqnarray}

However, the component of the curvature ${\Omega_{\pm\tilde a}}^{\mp\tilde b}$ remains unfixed by the BPS
equations, and it is determined only by the equations of motion. This is not surprising, corresponding
${\Omega_{\pm\tilde a}}^{\mp\tilde b}$ to the light-like deformation. We would like to comment that, at this
stage, the integrability condition we computed applies to null-deformation of {\it any} domain wall, curved or
flat.

As a consequence of GIC, we get the following equation for the ansatz \ref{metric}. The condition (\ref{Opmpm})
 gives
\begin{equation}
\frac {1}{\beta^2} \left(\frac{\partial_-^2 H}{k^2} - \left(\frac{\dot \beta}{\beta}\right)^2\right) = - g^2
W^2\label{1a},
\end{equation}
together with
\begin{eqnarray}
& &\partial_i\partial_-H =0,\label{1b}\\
& & \frac{\dot\beta^\prime- 2 \dot \beta\beta^\prime}{\beta^2} + \frac{\partial_-\dot H}{k^2}= - 3 g^2 \left(v\,
\partial^X W \partial_X W + w\, \frac 1{\gamma^2} \partial^x W \partial_x W \right) .\label{1c}
\end{eqnarray}
The condition (\ref{Opma}) provides
\begin{equation}
- \left[ \frac{\ddot \beta \beta -\dot \beta^2}{\beta^4}\delta_{\tilde a}^5 +
 \left(\frac{\dot \beta}{\beta^2}\right)^2 \right] =
g^2 \left[ 3 \left(\partial^X W \partial_X W + \frac 1{\gamma^2} \partial^x W \partial_x W \right)
\delta_{\tilde a}^5 - W^2\right],\label{2a}
\end{equation}
that together with (\ref{1a}) and  $\dot W = \sqrt{\frac 32} Q^s \dot \varphi^\Lambda D_\Lambda P^s = -3 g\gamma
\beta \left(\partial^X W \partial_X W + \frac 1{\gamma^2} \partial^x W \partial_x W \right) $
 implies
\begin{equation}
\frac{\dot \beta}{\beta^2} = g\gamma W, \ \ \ \ \ \ \gamma^2=1, \label{2b}
\end{equation}
and
\begin{equation}
\partial^2_- H = 0.\label{2c}
\end{equation}
In comparison with the DW case, we observe how the radial dependence of the warp-factor is still controlled by
the superpotential, with the difference that now $W$ can be also a function of $x^+$. As well, the relation
$\gamma^2=1$ indicating the ``flatness" of the wall is maintained: actually this is an input we put in
(\ref{metric}), focusing as announced on deformation of the (flat) DW metric (\ref{mconf}). The condition
(\ref{Ompa}) brings to
\begin{eqnarray}
& & \frac{\dot\beta^\prime- 2 \dot \beta\beta^\prime}{\beta^2} = - 3 g^2 \left(v\, \partial^X W \partial_X W +
w\, \frac 1{\gamma^2} \partial^x W \partial_x W \right),\\ \label{3a} & & {\partial_-\dot H} = 0. \label{3b}
\end{eqnarray}
The condition (\ref{Oab}) does not furnish any new relation. The independent equations for the ansatz are
summarized in the main text, eq. (\ref{dbeta}-\ref{H}).  The additional equation necessary to determine $H$ in
terms of the geometric quantities $W$, $v$, $w$ and $u^\Lambda$ will come from the equations of motion, eq.
(\ref{Rpmpm}).

 The last integrability condition to be considered comes from the scalar fields. Indeed, being now functions of $r$ and
 $x^+$, is necessary to check that $\partial_{[r}\partial_{x^+]} \varphi^\Lambda = 0$. The explicit expression is given
 in section \ref{null}, eq. (\ref{intsca}).

\section{Equations of motion} \label{eom}

The equations of motion of the lagrangian (\ref{lagra}) (taking in account also the terms containing the gauge
field that are zero for the configurations we study) for the metric, the gauge field and the scalars are,
respectively
\begin{eqnarray}
&&-R_{\mu\nu} +a_{IJ} F^I_{\mu a} F_\nu^{Ja} +g_{XY} D_\mu q^X D_\nu q^Y +g_{x y} D_\mu \phi^x D_\nu \phi^y
-\frac 16 |F|^2 g_{\mu \nu} +\frac 23 g^2\VV g_{\mu \nu} =0, \cr &&\label{einstein_v}\\
& & \nabla_a (a_{IK} F^{Kae}) +\frac 1{2 \sqrt 6} C_{IJK} \epsilon^{abcde} F^J_{ab} F^K_{cd} -g K^X_I D^e q^Y
g_{XY}=0, \label{maxwell_v}\\
 & & \hat{D}_\mu( D^\mu q^W) +gA^{\mu I} D_\mu K_I^W =  g^2 g^{WX} \partial_X
\VV, \label{qq_v} \\
& &  \hat{D}_\mu( D^\mu \phi^x) +gA^{\mu I} D_\mu K_I^x = g^2 g^{x y}
\partial_y \VV +\frac 14 g^{x y} \partial_y a_{IJ} F^I_{\mu \nu}
F^{J\mu \nu}, \label{fifi_v}
\end{eqnarray}
%Here $\DD$ is the covariant derivative with respect to the spin connection and
where $\hat{D}$ is a totally covariant derivative, ie with respect to all the indices, explicitly
\begin{eqnarray}
\hat{D}_\mu D^\mu \varphi^\Lambda =\nabla_\mu D^\mu \varphi^\Lambda +\Gamma^\Lambda_{\Sigma\Theta} D_\mu
\varphi^\Sigma D^\mu \varphi^\Theta \nonumber.
\end{eqnarray}

Specializing them to our uncharged configurations we get for the metric (we use the unifying notation for the
scalars)
\begin{equation}
R_{ab} = g_{\Lambda\Sigma} \partial_a\varphi^\Lambda \partial_b\varphi^\Sigma + \frac 23 g^2 {\cal V} \eta_{ab}.
\end{equation}
This identity can be easily checked for the component of the Ricci tensor following by the integrability
condition (\ref{intgranddw}) and the correspondent BPS values of the kinetic term of the scalars (\ref{SIC}) and
of the potential (\ref{scalarpotential}). Such result can be obtained applying the general result of
\cite{Celi:2003qk}. The $(\pm,\pm)$- component gives instead a new equation:
\begin{equation}
R_{\pm\pm} = -9 g^2 \left(v\, \partial^X W \partial_X W + w\, \partial^x W \partial_x W \right) - u^\Lambda
u_\Lambda. \label{Rpmpm}
\end{equation}
Making the comparison with the metric ansatz (\ref{metric}) one finds the constraint (\ref{Rpm}).

Although we are considering uncharged configuration (\ref{maxwell_v}) is not trivial. Indeed it entails
$K^X\partial_\mu q^Y g_{XY} = 0$. The main consequence of such condition is that the Hyperini equation becomes
of the same form of the gaugini equation. This fact may be seen as the deepest reason why the democratic
treatment of the scalars applies in the contest of DW solutions.

Regarding the e.o.m for the scalars, we observe that it reduces to the equation for the DW. Explicitly, we have
\begin{equation}
\nabla_\mu(\partial^\mu\varphi^\Lambda) + \Gamma_{\Omega\Sigma}^\Lambda \partial_\mu \varphi^\Omega \partial^\mu
\varphi^\Sigma = g^2 \partial^\Lambda {\cal V},\label{emsca}
\end{equation}
where the first term on the l.h.s. can be written as $\nabla_\mu(\partial^\mu\varphi^\Lambda) =
\partial_\mu\partial^\mu \varphi^\Lambda + (\partial_\mu\ln \sqrt{-g}) \partial^\mu \varphi^\Lambda$.
 It is immediate to verify that only the term with $\mu = r$ survives because NDDWs are cyclic in $x^-$ and $g^{\mu\nu}$
 is off-diagonal in $x^+$. Hence, (\ref{emsca}) reduces to scalar e.o.m. of the DW and the same manipulations
 hold, as the relation between $\partial_r \varphi^\Lambda$, $\beta$ and $W$ is unchanged in the deformed case.

\section{Parametrization of the two-dimensional projective quaternionic space}\label{app:paramCosets}

We shall consider the quaternionic-K{\"a}hler manifold of quaternionic dimension 2:
\begin{equation}
  \frac{\Sp(2,1)}{\Sp(2)\times \Sp(1)}\simeq
\frac{\USp(4,2)}{\USp(4)\times \USp(2)}.
 \label{Coset2}
\end{equation}
The algebra of the isometry group, $\fsp(2,1)$ can be defined as the set of matrices over the quaternions
$\mathbb{H}$ that preserve a metric of signature $(+,+,-)$. We take this metric in the form
\begin{equation}
  \mu =\begin{pmatrix} & & 1 \\
& 1 & \\
 1 & & \end{pmatrix},
 \label{mumetricSp}
\end{equation}
where each entry is a quaternion, or $2\times 2$ complex matrix. The elements  $M$ of $\fsp(2,1)$ are those
$3\times 3$ matrices with entries in $\mathbb{H}$ that satisfy
\begin{equation}
  \mu M^\dagger \mu = -M.
 \label{defnsp21}
\end{equation}

The general form of an element of $\fsp(2,1)$ is then
\begin{equation} M=
\begin{pmatrix}
a & \ft12(\bar e+\bar f)  & -\ft12(\vec b+\vec{c}) \\[1mm]
\ft12(e-f) & \vec p & -\ft12( e+f)  \\[1mm]
\ft12(\vec{b}-\vec c)  & \ft12(\bar f-\bar e) & -\bar a
\end{pmatrix},
\end{equation}
where $a=a_0+\vec a$, $e=e_0+\vec e$  and $f=f_0+\vec f$ are generic quaternions  and $\vec c$, $\vec b$ and
$\vec p$ are pure anti-Hermitian quaternions (with vanishing Hermitian part).\footnote{The identification
$\fsp(2,1)\simeq \fusp(4,2)$ is obtained once we take the matrices  $ -\ii \vec \sigma $ for the imaginary
quaternions.}

The Lie  algebra of $\fsp(2,1)$ can be split into  a compact (anti-Hermitian) and non-compact  (Hermitian) part
:
\begin{equation} M_{H}=
 \begin{pmatrix}
\vec a & \ft12\bar f & -\ft12\vec c \\[1mm]
-\ft12 f & \vec p & -\ft12 f  \\[1mm]
-\ft12\vec c &  \ft12\bar f & \vec a
\end{pmatrix}
, \quad M_{G/H}=
 \begin{pmatrix}
 a_0 \unity  & \ft12\bar e & -\ft12\vec b \\
\ft12 e & 0  & -\ft12 e  \\ \ft12\vec b &  -\ft12\bar e & - a_0\unity
\end{pmatrix}.
\end{equation}
The $H$ part of the generator can be decomposed into its subalgebras\footnote{It is related to the previous
expression of $M_{H}$ by  taking $\vec u=\ft12\vec a+\ft14\vec c$  and $\vec v=\ft12\vec a-\ft14\vec c$.} :
\begin{equation}  M_{\fsu( 2)}=
\begin{pmatrix}
\vec u  & 0 & -\vec u \\
0 & 0  & 0  \\
-\vec u &  0  &  \vec u
\end{pmatrix}
,\quad M_{\fsp(2)}=
\begin{pmatrix}
\vec v  & \ft12\bar f & \vec v \\[1mm]
-\ft12 f & \vec p  & - \ft12 f  \\[1mm]
\vec v  &  \ft12\bar f  & \vec v
\end{pmatrix}.
\end{equation}
$ M_{\fsp(1)}$ commutes with $M_{\fsp(2)}$ and the latter contains two commuting $\fsu(2)$ parameterized by
$\vec p$ and $\vec v$:
\begin{equation}
M_{\fsu(2)\oplus \fsu(2)\subset \fsp(2)}=
\begin{pmatrix}
\vec v  & 0 & \vec v \\
0       & \vec p       & 0         \\
\vec v  &  0  & \vec v
\end{pmatrix}.
\end{equation}
We see that the compact subalgebra of $\fsp(2,1)$ contains  three commuting $\fsu(2)$.
$M_{\fsu(2)}\subset\fsp(1)$ corresponds to the R-symmetry whereas the $\fsu(2)_{\vec p}\subset \fsp(2) $
contains the compact $\rU(1)$ for the string.

The solvable gauge of the coset manifold is obtained  by adding to $M_{G/H}$ an element of $M_H$ (with $\vec
c=\vec b$, $f=e$ and $\vec a=\vec p=0$) so that the result is an upper triangular matrix:
\begin{equation}
M_{\rm Solvable}=
 \begin{pmatrix}
 a_0 \unity & \bar e & -\vec b \\
0 & 0  & - e  \\
0 &  0 & - a_0 \unity
\end{pmatrix}.
\end{equation}

\subsection{Solvable coordinates and metric of $\frac{\Sp(2,1)}{\Sp(2)\Sp(1)}$}
\label{ss:coordSp21}

 We parametrize the coset elements by
\begin{equation}
L=\rme^N \cdot \rme^H,
\end{equation}
where
\begin{equation} N=N_e+N_b=
\underbrace{
\begin{pmatrix}
0 & \bar e & 0\\
0 & 0 & - e \\
0 & 0 & 0
\end{pmatrix}
}_{N_e} + \underbrace{
\begin{pmatrix}
0 & 0 & -\vec b \\
0 & 0 & 0 \\
0 & 0 & 0
\end{pmatrix}
}_{N_b} \, ,\quad H= \frac{1}{2}
\begin{pmatrix}
h \unity  & 0 & 0 \\
0 & 0 & 0 \\
0 & 0 & -h \unity
\end{pmatrix}.
\end{equation}
The coordinates $q^X$ are thus the real $h$, the 3 real coordinates of $\vec{b}$ and the 4 real parts of the
quaternion $e$. This leads to
\begin{equation} L=
 \begin{pmatrix}
\rme^{\frac{1}{2}h} \unity & \bar e & -\rme^{-\frac{1}{2}h}(\vec b +\frac{\bar e e }{2}) \\
0 & \unity & - \rme^{-\frac{1}{2}h} e \\
0 & 0 & \rme^{-\frac{1}{2}h}\unity
\end{pmatrix}.
\end{equation}
This leads to the algebra element
\begin{equation} L^{-1} \rmd L=
\begin{pmatrix}
\frac{B_0}{2} & \frac{\bar E}{\sqrt{2}} & -\vec B \\
0 & 0 & -\frac{E}{\sqrt{2}} \\
0 & 0 & -\frac{B_0}{2}
\end{pmatrix},
\end{equation}
where
\begin{equation}
B=B_0\unity+\vec B= \rmd h  \unity +\rme^{-h}\left[\rmd\vec b -\ft{1}{2}(\bar e \rmd e-\rmd \bar e e )\right]\,
,\quad  E= \sqrt{2}\, \rme^{-\frac{1}{2}h} \rmd e, \label{defBE}
\end{equation}
or in real components
\begin{equation}
B_0=\rmd h,\qquad B^r=\rme^{-h}\left({ \rmd b^r+  e^r \rmd e^0-e^0 \rmd e^r- \varepsilon^{rst} e^s \rmd e^t
}\right).
\end{equation}
The algebra element can be split in the coset part and the part in $H$. The first one is the Hermitian part:
\begin{equation}
(L^{-1} \rmd L)_{G/H}= \frac{1}{2}
\begin{pmatrix}
B_0 & \frac{ \bar E}{\sqrt{2}} & -\vec B \\
\frac{ E}{\sqrt{2}} & 0 & -\frac{ E}{\sqrt{2}} \\
\vec B & -\frac{\bar E}{\sqrt{2}} & - B_0.
\end{pmatrix}.
\end{equation}
The part in $H$ is the anti-Hermitian part, which can be split in the $\fsp(1)$ and $\fsp(2)$ part:
\begin{eqnarray}
(L^{-1} \rmd L)_{H}&=& \frac{1}{2}
\begin{pmatrix}
0 & \frac{\bar  E}{\sqrt{2}} & -\vec B \\
-\frac{E}{\sqrt{2}} & 0 & -\frac{E}{\sqrt{2}} \\
-\vec B & \frac{\bar E}{\sqrt{2}} & 0
\end{pmatrix}
=(L^{-1} \rmd L)_{\fsp(1)}+(L^{-1} \rmd L)_{\fsp(2)},
\nonumber\\
&&(L^{-1} \rmd L)_{\fsp(1)}= \frac{1}{4}
\begin{pmatrix}
 \vec B & 0 & -\vec B \\
0 & 0 & 0 \\
-\vec B & 0 & \vec B
\end{pmatrix}, \nonumber\\
&&(L^{-1} \rmd L)_{\fsp(2)} =
\begin{pmatrix}
-\frac{1}{4}\vec B & \frac{\bar  E}{\sqrt{2}} &-\frac{1}{4} \vec B \\
-\frac{ E}{\sqrt{2}} & 0 & -\frac{E}{\sqrt{2}} \\
-\frac{1}{4}\vec B & \frac{\bar E}{\sqrt{2}} & -\frac{1}{4}\vec B
\end{pmatrix}.
\end{eqnarray}
The metric is defined as
\begin{equation}
\rmd s^2=g_{XY}\rmd q^X\rmd q^Y  =  \Tr \left[{ (L^{-1} \rmd L)_{G/H}\cdot (L^{-1} \rmd L)_{G/H} }\right] =
\ft{1}{2}\tr ( B\bar B+ E\bar E ), \label{ds2def}
\end{equation}
where $\Tr$ stands for a trace over the $6\times 6$ matrix and $\tr$ for a trace over the $2\times 2$ matrix. We
will comment on the normalization of this metric below. Its value is
\begin{equation}
\rmd s^2= (\rmd h)^2+ (B^1)^2+(B^2)^2+(B^3)^2 +2\rme^{-h}\left[{ (\rmd e^0)^2+(\rmd e^1)^2+(\rmd e^2)^2+(\rmd
e^3)^2 }\right]. \label{metricSp21}
\end{equation}
The vielbeins, as 1-forms and quaternions as explained above, can be taken to be
\begin{equation}
  f^1= \frac{1}{\sqrt{2}} B,\qquad f^2 =\frac{1}{\sqrt{2}} E.
 \label{f1f2}
\end{equation}
These lead to (\ref{ds2def}) and to the hypercomplex form ($\wedge $ symbols understood)
\begin{equation}
  \vec{J}=-\ft12 \left(\bar B\,  B+\bar E\,  E\right),\qquad \mbox{or}\qquad
  J^r=-B_0B^r -E_0 E^r -\ft12\varepsilon ^{rst}\left( B^sB^t
  +E^s E^t\right).
 \label{vecJmatrix}
\end{equation}
Using the differentials
\begin{eqnarray}
  &&\rmd B= -B_0  B-\ft12\bar E\, E, \qquad \rmd E=-\ft12B_0
  E,\nonumber\\
   &&\mbox{or}\qquad \rmd B^r= -B_0 B^r-E_0E^r-\ft12\varepsilon
  ^{rst} E^sE^t,
 \label{dBdE}
\end{eqnarray}
we obtain
\begin{equation}
  \rmd J^r +2\varepsilon ^{rst}\omega ^sJ^t=0,
 \label{dJomega}
\end{equation}
for
\begin{equation}
  \omega ^r=-\ft12 B^r.
 \label{valueomega}
\end{equation}
We find then that (\ref{defSU2cR}) is satisfied for $\nu =-1$. The value that we get here for $\nu $ depends on
the normalization of the metric. Multiplying the metric by an arbitrary $-\nu^{-1}$, would lead to
(\ref{defSU2cR}) with this arbitrary value of $\nu $. In the supergravity context, $\nu = -\kappa ^2$, where
$\kappa $ is the gravitational coupling constant, which we have put equal to~1.

\section{Adapted coordinates} \label{adapt}

In this section we present some insights on the SIC and the possible solutions it admits. The crucial ingredient
is the adoption of adapted coordinates, associated to an existing solution. This choice allows to emphasize the
physics of the solution, that is characterized by two dynamical scalars. As a result, the properties of the
possible solutions can be better understood, even without constructing them explicitly. However, one has to have
clear the price paid assuming a priori the existence of a solution. We will further comment on this point.

Generalizing the argument in \cite{Celi:2004st}, the metric of the moduli space on the two-dimensional
sub-manifold identified by a solution can be cast as
\begin{equation}
\left. g_{\Lambda\Sigma}\right|_{\text{sol}}= \left( \begin{array}{ccc}
g_{11} & g_{12} & 0 \\
g_{12} & g_{22} & 0 \\
0      &   0     &   g_{\hat{\Lambda}\hat{\Sigma}}
\end{array} \right), \label{adaptedmetric}
\end{equation}
where $\varphi^1$ and $\varphi^2$ represent the dynamical scalars while the others are constant. In this optic
the meaning of the SIC is more clear. The different solutions for a given $W$ coincides with the possible
embedding  of  a two-dimensional submanifolds ${\cal I}_{\text{sol}}$ admitting a metric of the form
(\ref{adaptedmetric}).

To be concrete, let us study this problem in presence of $u$-deformation only. In this special case, as observed
in section \ref{anal}, the $r$ and $x^+$ dependence ``decouple", being $W=W(r)$ and $D=D(x^+)$. This further
simplifies our problem. $W=W(r)$ implies the existence of a preferred coordinate system in which $\varphi^1 =
W$. This parametrization is well-defined until we are out of the critical points of superpotential, i.e.
$\partial_\Lambda W \neq 0$. The clear advantage of this coordinate choice is that $u^\Lambda$ lies in the
direction $2$ and $\varphi^1$ depends only on $r$, $u^1=(\varphi^1)^\prime = 0$. Because of this (\ref{intsca})
gives
\begin{eqnarray}
&& D g^{11} = -u^2 \partial_2(g^{11}),\label{1}\\
&& D g^{12} + \frac 13 \varphi^1 u^2 = (g^{11}\partial_1 + g^{12} \partial_2) u^2 - u^2 \partial_2
g^{12}.\label{2}
\end{eqnarray}
These equations can be formally integrated in terms of $\varphi^1$ and $\varphi^2$ taking in account that
$(\varphi^1)^\prime = 0$ together with (\ref{1}) implies
\begin{equation}
\frac D{u^2} = \partial_2 \ln \beta = - \partial_2 \ln g^{11}.\label{3}
\end{equation}
This leads to a warp-factor of the form $\beta = \frac{F(\varphi^1)}{g^{11}}$; the function $F$ can be computed
using (\ref{1}) or equivalently (\ref{dbeta}). The resulting $\beta(\varphi^1,\varphi^2)$ is
\begin{equation}
\beta = e^{-\frac 13 \left[ \int \left(g^{12}\partial_2(\frac 1 {g^{11}}) - \frac {\varphi^1}{g^{11}}\right)
\rmd \varphi^1\right]}. \end{equation} Equivalently, a formal expression for $u^2$ can be obtained from
(\ref{2}) solving \begin{equation} (g^{11}\partial_1 + g^{12}\partial^{12}) \ln u^2 = \partial_2 g^{12} - g^{12}
\partial_2 \ln g^{11} + \varphi^1.
\end{equation}
As in the explicit example of section \ref{sol} $u^\Lambda$ is not completely determined as a function of the
moduli space. In terms of the previous equations, the spacetime parametrization of the scalars is
\begin{eqnarray}
& &\dot \varphi^1 = -3g \gamma \beta g^{11} = -3g\gamma F(\varphi_1),\\
& &\dot \varphi^2 = -3g \gamma \beta g^{12} = -3g\gamma \frac {g^{12}}{g^{11}} F(\varphi_1),\\
& &{\varphi^2}^\prime = \beta u^2 = \frac {u^2}{g^{11}} F(\varphi_1).
\end{eqnarray}

The above equations deserve some comments. As we stress at the beginning they can be interpret only as a formal
solution. Indeed, we start assuming that the solution exists: this implies that the coefficients of the
effective two dimensional metric are non generic, in order to guarantee the existence of the solution. This can
be understood considering $F(\varphi_1)$: $\partial_2 F = 0$ ends up in an integrability condition on the such
coefficients. Hence the integrability requirements of SIC are just rewritten in a different way.

Keeping in mind this caveat, the adapted coordinates are still an useful tool. For example, they make clear that
we may have a solution with non trivial $D$ even if the two dimensional metric is diagonal, i.e. $g^{12} = 0$.
As (\ref{3}) shows, the crucial condition is $\partial_2 g^{11} \neq 0$. Unfortunately, this does not occur for
the examples considered in section \ref{sol}.

%%%%%%%%%%%%%%%%%%%%%%%%%%%%%%%%%%%%%%%%%%%%%%%%%%%%%%%
%\bibliography{nddw}
%Included for WinEdt Gather Purpose (do not remove the comment line below:
             %input "C:\localtexmf\bibtex\bib\*.bib"
%\bibliographystyle{toine}

\providecommand{\href}[2]{#2}\begingroup\raggedright\endgroup

\end{document}